# Phenomenon-first problem formulation with language models in metal–organic frameworks

Jihan Kim

Department of Chemical and Biomolecular Engineering, Korea Advanced Institute of Science and Technology (KAIST), Daejeon, Republic of Korea

Correspondence: jihankim@kaist.ac.kr

## Abstract

Here we introduce a phenomenon-first framework for materials discovery in which an LLM receives separate descriptions of guest physics, host capabilities and operating context and proposes an unusual qualitative behavior with its mechanism, distinguishing signature and falsifier. Across 80 guest–host combinations in metal–organic frameworks, the model generated 57 proposals that were frozen before evaluation or materials selection; assessment identified nine computationally tractable candidates. In a repeat and guest-only baseline, structured formulation reproducibly broadened mechanistic exploration: the original and repeat runs each covered 21 broad mechanism families versus eight for guest-only formulation despite fewer proposals. From the original run, MC-010 yielded reproducible hydration-driven reorientation of retained ethanol in experimentally reported MOFs, while MC-015 yielded a quantum-corrected reversal of $H_2$ adsorption-region preference relative to the matched classical description. These results show how structured LLM reasoning can move upstream from executing supplied objectives toward prospectively formulating qualitative materials phenomena.

Automation and large language model (LLM) agents are expanding research tasks in chemistry and materials science. Self-driving laboratories connect experimental design, synthesis and characterization,[1–3] while tool-using systems such as ChemCrow and Coscientist coordinate instruments and software.[4,5] Computational agents now automate broad atomistic workflows, including AtomAgents, MDCrow and El Agente.[6–13] Much of this work centers on executing a supplied objective through planning, tool use and analysis.

A less explored question is whether such systems can help specify the objective itself. Robot scientists generated and tested biological hypotheses,[14] AI Co-Scientist generates and refines hypotheses around scientist-defined goals,[15] and the AI Scientist links idea generation with computational experimentation.[16] SciMON and SciAgents use retrieval, knowledge graphs and multi-agent reasoning to generate hypotheses.[17,18] Related autonomous MOF work launched 16 separately initialized campaigns from the same methane-storage objective, revealing divergent strategies but convergent material frontiers.[19] A particularly close precedent is peptide work in which an agent proposed and computationally investigated previously unreported qualitative relationships.[20] These studies show language models can contribute to formulation as well as execution. However, to our knowledge, no previous study has prospectively formulated an unspecified qualitative phenomenon and then identified and computationally tested realizations in inorganic or hybrid materials. Here the physical behavior itself, rather than a route to a supplied target, becomes the object of the search.

Metal–organic frameworks (MOFs) provide a natural setting in which to test this capability. Guest behavior emerges from coupling molecular properties such as size, electrostatics, hydrogen-bonding propensity and polarizability with host confinement, binding environments, connectivity, heterogeneity and structural degrees of freedom.[21–24] Temperature, pressure, loading and composition further shape these interactions. Established examples include cooperative $CO_2$ insertion in diamine-appended MOFs,[25] adsorption-induced structural transitions[26] and negative gas adsorption.[27] The chemical diversity of reported MOFs and availability of computation-ready structural databases also make it practical to search for hosts that satisfy the requirements of a prospectively formulated phenomenon.[28–35]

Here we investigate this phenomenon-first mode of scientific problem formulation. Rather than asking an LLM to brainstorm unrestricted research ideas or optimize a predefined material property, we provide separate descriptions of guest-side physics, host capabilities and operating context and ask the model to infer a non-obvious qualitative behavior that could emerge from their interaction. Each proposal must specify a physical mechanism, distinguishing observable, relevant conditions and host requirements, and a falsifier; abstention is permitted when the supplied ingredients do not support a distinctive phenomenon. We apply this procedure prospectively to 80 combinations of ten guest systems and eight host archetypes, freeze the outputs before literature auditing or material selection, and evaluate their physical coherence, relationship to prior work and computational realizability. Of nine near-term candidates retained after filtering, MC-010 and MC-015 were developed as physically distinct principal cases: hydration-driven reorientation of retained ethanol in real MOFs and quantum-confinement-driven redistribution of adsorbed hydrogen in a matched classical–semiclassical comparison. This design tests whether a model can formulate a falsifiable materials phenomenon before the material that realizes it is chosen.

## Results

### Constructing the inputs and generating prospective phenomena

The original discovery workflow is summarized in Fig. 1. GPT-6 Astra generated ten guest knowledge cards, eight host capability cards and application-specific operating contexts, all frozen before prospective phenomenon generation. Each later call received one guest card, one host card and the corresponding context, but no target phenomenon or specific material. Proposals were subsequently assessed for physical plausibility, relationship to prior literature and computational realizability.

GPT-6 Astra selected five unary and five binary guest systems spanning gases, molecular adsorbates and mixtures (Table 1). The cards summarized established properties relevant to confinement, including molecular geometry, electrostatics, polarizability and intermolecular interactions, and were explicitly restricted from proposing unusual behavior. Their authoring procedure is described in Supplementary Sections S1.4 and S2.1–S2.2; complete frozen card texts are retained in the study archive.

The model also generated eight host archetypes describing porous-material capabilities rather than named MOFs (Table 2), spanning confinement, binding sites, patterned surfaces, local and collective motion, pore heterogeneity, vibrations and external entrances. The cards described each capability and information needed later to identify a real material without assigning a preferred guest, occupancy order or response direction; a real MOF could satisfy more than one archetype.

Each guest–host pair additionally received an application and operating context generated before phenomenon generation. These specified the relevant feed and qualitative regime while leaving material-specific numerical conditions for later calculations. The complete context inventory and assignments are provided in Supplementary Sections S2.5–S2.7 and Tables S1–S2.

These components defined an exhaustive 80-cell matrix. Each cell was evaluated once in an independent GPT-6 Astra call using max reasoning. A non-abstaining response had to provide one qualitative phenomenon with its physical basis, distinguishing signature, relevant conditions, application relevance, falsifier and host requirements; ordinary improvements in uptake, binding or selectivity were insufficient. Calls had no access to other generated responses, literature searches, material structures or subsequent simulation results, and abstentions were not regenerated.

**From the original proposals to two principal case studies**

The 80 calls produced 57 proposals and 23 abstentions. Because the same guest physics could support different questions in different host environments, the outputs spanned distinct mechanisms; for example, $H_2$ generated tunneling-supported exchange, inverse thermal partitioning and mode-hybridization hypotheses in different host archetypes. Selected outputs are given in Supplementary Table S3, the complete outcome and family map in Table S2, and full unedited responses in the study archive.

Several proposals expressed the same underlying causal mechanism in different chemical settings. Grouping yielded 28 families, which were assessed for physical coherence, relation to prior literature and computational feasibility. The literature audit classified 14 as close precedent, 13 as related but distinct and one as unresolved. The 14 not classified as close precedent were then assessed for execution: two received tier A, seven tier B and five tier C. The nine tier-A/B mechanisms were retained as near-term candidates and translated into host requirements before specific MOFs were selected. Full grouping criteria and records are provided in Supplementary Sections S4.1–S4.6.

The nine retained candidates were MC-004, MC-006, MC-010, MC-015, MC-018, MC-019, MC-020, MC-022 and MC-025. Five were developed far enough to reach a practical decision: MC-004 was deprioritized after its novelty distinction narrowed, MC-020 was deferred because its decisive test became disproportionately demanding, and MC-025 did not yield a defensible host realization. MC-010 and MC-015 produced positive, physically complementary outcomes and became the principal case studies. The other four remained prospective but were not pursued further; their advancement histories are given in Supplementary Section S4.7 and Table S4.

**Repeat generation and comparison with guest-only formulation**

To examine consistency across generation runs, we repeated all 80 calls with the same frozen inputs and explicit settings. The repeat produced 63 proposals and 17 abstentions, with proposal-versus-abstention decisions matching the original for 74 of 80 inputs (92.5%). All 57 inputs that originally yielded a proposal again did so, while six of 23 original abstentions produced proposals in the repeat. Individual texts were not deterministic reproductions, but mechanism-level recurrence was evident: MC-015 was independently regenerated from the same $H_2$–heterogeneous-pore input, whereas MC-010 did not recur among the 63 repeat proposals. Details are provided in Supplementary Section S3.6.

We next compared mechanistic breadth across the original structured run, repeat structured run and an 80-call guest-only baseline. Each baseline call retained the corresponding guest card and operating context but omitted the host card, leaving the model to infer required host characteristics. The three runs each comprised 80 calls and produced 57, 63 and 80 proposals, respectively. A fresh GPT-6 Astra evaluation jointly grouped all 200 anonymized proposals under a common causal-mechanism rubric, with run labels and historical assignments withheld (Supplementary Sections S3.7 and S4.8).

Under this shared taxonomy, the original and repeat structured runs each covered 21 broad mechanism families, compared with eight for guest-only formulation; subtype counts were 39, 41 and 21, respectively (Fig. 2a). Both structured runs covered more broad families than the guest-only baseline for every guest system (Fig. 2b). Sixteen broad families occurred in both structured runs, including 13 absent from guest-only. The guest-only outputs were also more concentrated: its two largest families contained 53 of 80 proposals (66.3%), compared with 15 of 57 (26.3%) and 18 of 63 (28.6%) in the original and repeat structured runs (Fig. 2c). Full counts are given in Supplementary Section S4.9 and Tables S9–S11.

Thus, under equal generation budgets, supplying and systematically varying host capabilities reproducibly broadened the mechanistic space explored despite yielding fewer proposals. The comparison tests the combined effect of supplying host information and systematically varying it across guest–host combinations; it does not isolate those contributions or establish greater novelty, physical validity or computational usefulness. We next return to the original discovery run and examine MC-010 and MC-015 as the two principal case studies.

## Case study 1. MC-010 hydration-driven inversion of retained ethanol

### *Generated hypothesis and operational test*

MC-010 arose from the ethanol/water guest card paired with a spatially patterned hydrogen-bonding surface. The generated hypothesis proposed that hydration could reorganize the orientation of ethanol already retained within a pore: dry ethanol would favor an OH-to-framework anchored configuration, whereas hydration could stabilize a flipped retained state with the hydroxyl group directed toward water and the ethyl group accommodated near a less-polar framework region. The defining prediction was therefore a change in retained-ethanol orientation, not simply increased water uptake or ethanol displacement. Water clustering in MOFs and the competing roles of hydroxyl hydrogen bonding and alkyl dispersion in alcohol adsorption are well established.[36–40]

We described the local configurations using three operational states: A, the dry anchored retained state; B, the flipped but retained state in essentially the same pocket; and C, competing configurations including alternative orientations, relocation and displacement. At finite temperature, the operational test is a substantial and reproducible redistribution from an A-containing dry ensemble toward B rather than nearly pure A and B populations. A stricter equilibrium realization would require B to lie below both A and the competing C ensemble in free energy.

### *Mechanistic stress tests define the design window*

A framework-free ethanol–water calculation first tested whether hydration could supply a sufficiently large orientational bias. At 300 K, the hydroxyl-facing preference increased from approximately 9.0 kJ $mol^{-1}$ with one water to 16.9 kJ $mol^{-1}$ with twenty waters; a twelve-water separation sensitivity gave approximately 12.5–18.0 kJ $mol^{-1}$. These calculations established that water can provide an orientational driving force large enough to compete with differences among physisorbed configurations.

Canonical hosts then exposed the complementary material requirements. Zn-MOF-74 anchored ethanol too strongly for retained reorientation, UiO-66-$NH_2$ allowed a relatively small A–B energetic gap but did not provide a persistent retained B basin, and pristine MOF-801 supplied confinement but not the chemical asymmetry needed to distinguish A from B. A deliberately symmetry-broken MOF-801-derived model showed that these requirements can be satisfied simultaneously: for a variant with a dry A–B penalty near 2.1 kJ $mol^{-1}$, one to six local waters produced B populations of approximately 59–75%, whereas larger dry penalties increasingly favored competing configurations. The combined tests therefore defined the design window as moderate dry anchoring, an independently retained flipped basin, preferential hydration of its exposed hydroxyl group and sufficient confinement to suppress alternative configurations. Detailed intermediate results are provided in Supplementary Sections S5.3–S5.5.

***Real MOFs exhibit reproducible hydration-driven reorientation***

Having established the design window in the symmetry-broken MOF-801-derived model, we screened 46 constructed MOF-801 variants and a pristine control under dry conditions. None passed the combined qualification criteria for hydration testing, so we broadened the search to the working CoRE MOF 2024 corpus. Static screening was completed for 10,700 retained entries, and promising pockets underwent matched A- and B-start validation. The completed campaign examined 291 leads using eight matched chains per candidate, reusing verified trajectories where appropriate.

The two most informative real-MOF examples are summarized in Fig. 3. For 2019[Zn][sqp]3[FSR]1, the dry ensemble contained approximately 45.2% A, 5.2% B and 49.6% C, whereas four local water molecules shifted the sampled distribution to approximately 0% A, 97.2% B and 2.8% C (Fig. 3a–c). All four independent hydrated trajectories strongly populated B (approximately 93–99%) from both A- and B-start configurations. Thus the dry state need not be pure A: the key result is the large, reproducible redistribution from a substantial anchored-A population to the specific retained B orientation without loss of ethanol from the local pocket. For 2012[Cu][rtl]3[FSR]2, the dry ensemble contained approximately 46.2% A and 1.9% B, whereas hydration increased B to approximately 48.3% (Fig. 3d–f).

Matched validation, rather than a single favorable trajectory, provides the main evidentiary weight. Each promoted lead was tested dry and hydrated from both A- and B-like starts with two independent seeds per start. For 2019[Zn][sqp]3[FSR]1, all four hydrated chains converged on the same B-rich behavior while ethanol remained localized, arguing against initialization artifacts or simple displacement. The independent 2012[Cu][rtl]3[FSR]2 framework reproduced the same redistribution direction with less complete enrichment, showing that the response was not confined to the deliberately favorable development model.

Across the full 291-lead matched-validation campaign, 8 leads were classified as reproducible A-to-B-like, 151 as strong partial phenomena, 79 as bistable or initialization-dependent, and 53 as not reproduced or showing no useful signal (Supplementary Section S5.7). For 2019[Zn][sqp]3[FSR]1, single-start checks at six and eight waters did not reproduce the strong four-water B state, indicating a finite hydration window rather than a monotonic increase of B occupancy with water loading (Supplementary Section S5.8).

Taken together, MC-010 provides a successful local realization of the prospectively generated phenomenon: hydration can reproducibly reorganize retained ethanol toward the predicted flipped state in real MOFs. The reported trajectory fractions are finite-run populations rather than precise equilibrium probabilities, so a stronger claim of a unique free-energy-basin switch would require dedicated free-energy calculations. The original proposal also included a collective extension in which shared water bridges couple multiple retained ethanol molecules; that many-body component was not tested here.

### Case study 2. MC-015 quantum-confinement-driven reversal of hydrogen regional preference

#### *Generated prediction and discriminating test*

MC-015 originated from pairing the molecular-hydrogen guest card with the host archetype describing connected, spatially heterogeneous pore environments. The supplied input stated only that low-temperature quantum effects can matter for $H_2$ and that the host contains distinguishable connected regions; it did not assign an adsorption-region ordering, a deep/sharp versus shallow/soft competition, a crossover temperature or a specific MOF. The generated response introduced the central physical relationship: a classically deeper region can confine $H_2$ more sharply than a shallower competing region, so the larger quantum-confinement penalty of the deeper region can reverse their statistical preference on cooling. The provenance boundary between supplied information, generated additions and later implementation is detailed in Supplementary Table S5.

The response also specified the discriminating observable before material selection. At low loading and without host transformation, the normalized A/B population ratio should decrease on cooling under a quantum-corrected description while a matched classical calculation using the same host and interaction landscape should move in the opposite direction. Loss of the inverse redistribution after sufficient equilibration was identified as a falsifier. Thus the prospective target was not simply a quantum correction to total $H_2$ uptake, but a reversal of relative population between two identifiable adsorption regions.

### *Computational implementation and corpus screen*

To obtain a tractable test, we used rigid frameworks, neutral single-site Buch $H_2$[41], UFF framework interactions[42] and a quadratic Feynman–Hibbs (FH) correction.[43,44] Relative dilute regional populations were obtained by integrating Boltzmann weights over fixed spatial basin masks and dividing each regional-family weight by its symmetry multiplicity before forming the A/B ratio. Classical and FH calculations used the same basin definitions and underlying interactions. Numerical implementation, multiplicity treatment and Widom validation are described in Supplementary Sections S6.2–S6.4. Prior work has established the importance of nuclear quantum effects for cryogenic $H_2$ adsorption in MOFs.[43–45]

The corpus screen examined 12,093 CoRE entries representing 8,887 coordinate-unique records. Energy fields were generated for 456 structures and 174 yielded usable regional assignments. The screen identified 53 candidate crossover pairs in 26 structures, including seven stronger candidates in four structures; three leads were characterized further. Two checked 2019 records share the same underlying host coordinates and differ by four retained DMF molecules, while the 2014 record provides a distinct framework. These counts describe the implemented screening pipeline rather than phenomenon prevalence; full accounting is given in Supplementary Table S6.

***Opposite classical and quantum-corrected temperature responses***

The primary checked lead was 2019[Zn][nan]3[FSR]24 (analysis ID m157), with regions R2 and R7 denoted A and B. Figure 4a,b compares the local classical and FH effective-potential landscapes for these competing regions at 25 K. At the marked representative positions, the local energy difference, $\Delta U_{B-A} = U_B - U_A$, changes from +0.50 kJ mol$^{-1}$ classically to −0.22 kJ mol$^{-1}$ with the FH correction, showing that the local ordering of the marked A and B positions reverses. The population-level consequence is shown in Fig. 4c. The FH A/B regional ratio decreased from 1.520 at 100 K to 0.595 at 25 K and crossed unity near 35 K, whereas the corresponding classical ratio increased from 1.660 to 4.541. Thus cooling strengthened the classical preference for A but reversed the FH-corrected preference toward B. The ratios in Fig. 4c are obtained from integrated regional statistical weights rather than from the two marked point energies in Fig. 4a,b; for this R2/R7 pair, both site families have symmetry multiplicity four, so the multiplicity factors cancel. The corresponding population fractions and normalization details are reported in Supplementary Table S7.

***Scope of the MC-015 realization***

The primary result provides model-level support for the prospectively generated regional-preference signature, but not a universal local-energy criterion. Across checked cases, the FH penalty at the classical minimum of A was larger than at B, yet non-switchers showed that the initial classical preference and integrated basin weight also matter. Regional assignments are approximate, FH is semiclassical at low temperature, and the configurational integral predicts equilibrium population ratios rather than exchange kinetics. We therefore do not claim experimental switching or explicit quantum validation. Supporting leads and sensitivity analyses are provided in Supplementary Sections S6.7–S6.9 and Table S8.

## Discussion

By following the generated questions into detailed computational work, this study examined how language models could help formulate useful directions for materials research. The value of each proposal became clearer as the requirements of a test were worked out. Some were set aside after close precedents were identified in the literature, while others required evidence beyond the scope of the study or could not be mapped to a defensible material realization. These outcomes helped determine which questions warranted further effort.

The repeat and guest-only comparisons provide a complementary view of the formulation stage. Repeating the original 80 calls reproduced the proposal-versus-abstention outcome for 74 of 80 inputs, while the common comparison across the original structured run, repeat structured run and guest-only baseline identified 21, 21 and 8 broad mechanism families, respectively. The guest-only outputs were also much more concentrated in a small number of mechanisms. These results suggest that supplying and systematically varying host capabilities broadens the range of questions explored, without establishing that the resulting ideas are necessarily more novel, physically correct or useful.

Among the proposals that advanced through this process, MC-010 and MC-015 illustrate two different levels of realization. MC-010 guided simulations that found reproducible hydration-dependent reorientation of retained ethanol in experimentally reported MOFs, whereas MC-015 provided a model-level demonstration in which a matched classical–Feynman–Hibbs comparison reversed regional $H_2$ preference in a manner consistent with the predicted competition between classical binding and quantum confinement. In both cases, the qualitative prediction and distinguishing signature were fixed before the downstream computational protocol was developed, allowing the calculations to test a physical relationship specified in advance.

These demonstrations nevertheless remained strongly researcher-directed. The researcher defined the scientific framing, guided construction of the guest–host representation, decided which proposals to advance, selected or approved the downstream methods and interpreted the resulting evidence. Developing meaningful tests also required exploratory calculations and simplified models to build physical intuition. The generated hypotheses were therefore developed and tested within a human-directed investigation rather than an end-to-end autonomous workflow.

This limitation also points to an important direction for follow-up work. For unfamiliar phenomena, agents would need to do more than execute predefined workflows: they would need to use exploratory calculations to refine tests, interpret partial results and decide when further computation is justified. With these limitations in mind, phenomenon-first reasoning offers a way to place question formulation alongside calculation and optimization in computational materials workflows, while recording generated hypotheses before material selection and method development keeps their contribution open to scrutiny.

## Methods

### Prospective generation and assessment

The guest–host–context representation, development exercises, exact prospective-generation instruction, request settings, mechanism grouping, scientific reviews, literature audit and executability assessment are documented in Supplementary Sections S1–S4. Briefly, ten guest knowledge cards and eight host-capability cards were frozen before prospective generation. Each of the 80 guest–host combinations was evaluated independently by GPT-6 Astra using one guest card, one host card and an operating context. The model returned one proposed qualitative phenomenon or ABSTAIN; no specific MOF, web search, retrieval or external tool access was supplied during these prospective calls. Non-abstaining outputs were subsequently grouped by causal mechanism, evaluated against prior literature and assessed for computational executability before host specification and researcher-directed advancement. These discovery-stage classifications and advancement decisions were retained separately from the comparative regrouping described below.

### Generation comparison and common mechanism grouping

To examine generation consistency, we repeated all 80 structured requests using the same frozen guest cards, host cards, operating contexts and explicit generation settings. An audit of the submitted request records verified that each repeat model-facing request body was identical to its corresponding original request. We additionally generated an 80-call guest-only baseline. Each guest-only input preserved the exact guest card and operating context for the corresponding slot while removing the complete host card, including its identifier and supplementary fields. The common generation instruction was minimally adapted to remove the assumption that host information had been supplied and to ask the model to infer the host characteristics required for its proposed phenomenon. No host-capability menu or source-case mapping was supplied. Some guest/context payloads therefore recurred across baseline slots and were retained as separate generation opportunities rather than deduplicated. Review of the saved contexts found no explicit disclosure of the omitted host archetypes.

The original structured, repeat structured and guest-only runs each comprised 80 generation calls. The repeat structured and guest-only calls used the same explicit generation settings as the original experiment: model identifier gpt-6-astra, max reasoning, medium text verbosity and a 65,536-token output ceiling; temperature, top_p and seed were not explicitly submitted. No tools, web access, retrieval, repository access, attachments or prior conversation history were supplied during generation. The guest-only condition therefore differed from the structured condition in the omission of supplied host information and the corresponding minimal instruction change. The same requested model alias and explicit settings do not establish an unchanged underlying serving snapshot across execution dates.

For the comparative mechanism analysis, the 57 proposals from the original structured run, 63 proposals from the repeat structured run and 80 proposals from the guest-only run were jointly regrouped under one common taxonomy. One fresh, isolated GPT-6 Astra evaluation received all 200 anonymized proposal texts together with a frozen causal-mechanism rubric. Broad families required a shared causal core, assessed from the causal sequence, necessary physical ingredients, distinguishing qualitative signature and falsifier. Subtypes retained meaningful variants within a broad causal family. Differences in guest identity or wording alone did not justify splitting, whereas differences that changed the essential physical explanation or decisive test required separation. No target family or subtype counts were supplied, and the evaluator was not asked to improve or repair proposals.

Run labels, source identifiers, historical family assignments and counts, literature and execution verdicts, shortlists and downstream simulation or manuscript outcomes were withheld from the evaluator. Administrative identifiers were removed while scientific proposal content was preserved. The resulting assignments were frozen before source identities were restored. This procedure therefore provided run-label blinding, although proposal content or writing style could still contain information from which provenance might be inferred. Mechanistic breadth was quantified as the number of broad families and subtypes represented in each 80-call run and within each guest system; family counts across guest systems are not additive because a family can occur for more than one guest. Seven grouping boundaries remained explicitly recorded as uncertain, and their alternative assignments were not adopted after unblinding. The comparative taxonomy was used only for the mechanistic-breadth analysis and did not replace the original discovery-stage 28-family classification. Complete protocols and tabulations for the repeat, guest-only and common-grouping analyses are provided in Supplementary Sections S3.6–S3.7 and S4.8–S4.9 and Tables S9–S11.

### MOF structures and pore information

A locally archived CoRE MOF 2024[33] snapshot supplied the 12,093-entry working corpus used by both campaigns. The MC-015 coordinate-based fingerprint grouped these entries into 8,887 records; local snapshot provenance and campaign-specific corpus accounting are given in Supplementary Sections S5.6 and S6.5. Framework electrostatics used PACMAN-predicted partial charges supplied with the working CoRE structures[46]. Existing Zeo++ calculations[34] supplied pore-limiting diameter, largest-cavity diameter, accessible volume and surface-related descriptors, which were reused where available. Exact structural provenance was established from CoRE mappings or original crystallographic sources for individually examined materials.

### MC-010 simulation and analysis

MC-010 used TraPPE-UA ethanol[47] and SPC/E water[48]. Framework Lennard-Jones interactions used DREIDING parameters for nonmetal framework atoms[49] and UFF parameters for metal atoms[42], together with PACMAN-predicted DDEC6 framework charges[46]. For the later host-containing calculations, including the real-MOF campaign, framework–ethanol and framework–water electrostatics used the periodic host Ewald potential. Guest–guest electrostatics in hydrated runs used direct full-range Coulomb interactions within the finite local cluster, without a pair cutoff, minimum-image wrapping or guest-image sums. Framework-free clusters also used direct guest–guest electrostatics; distinct guest-image treatments in the early Zn-MOF-74 calculations are documented in Supplementary Section S5.9. Three operational states were defined from saved ethanol trajectories: A, a localized framework-anchored orientation; B, a flipped but retained orientation in essentially the same local pocket; and C, the remaining alternative, relocated or displaced configurations. Reported A/B/C values are production-frame fractions and are treated as finite-run populations rather than equilibrium probabilities. Mechanistic development included framework-free ethanol–water calculations, canonical-host tests and a symmetry-broken MOF-801-derived model, all at 300 K. Simulation settings and parameter-file provenance are given in Supplementary Sections S5.1–S5.5 and S5.9.

Of the 12,093 working CoRE entries, 1,390 were excluded by descriptor criteria and three were technically deferred, leaving 10,700 entries that all completed static geometric/energetic screening. Selected pockets then underwent dry-chain and hydration calculations. The completed matched-validation campaign contained 291 leads, each represented by dry and hydrated A- and B-start trajectories with two independent seeds per start. The principal hydrated condition contained four local water molecules; new matched chains used 5,000 burn-in and 5,000 production sweeps, while previously verified trajectories were reused when they matched the frozen conditions. Production frames were classified with the frozen A/B/C definitions while localization, drift, seed agreement and mechanism-relevant contacts were retained as diagnostics. Structural panels use representative saved frames without coordinate optimization and do not imply recorded transition paths. Screening procedures, validation design and analysis definitions are provided in Supplementary Sections S5.6–S5.9.

**MC-015 model and regional integration**

MC-015 was tested in rigid frameworks using neutral single-site Buch $H_2$[41], UFF framework van der Waals interactions[42] and a quadratic Feynman–Hibbs (FH) correction[43,44]. Because the Buch $H_2$ site is neutral, the framework partial charges do not contribute to the $H_2$–framework interaction energy in this model. Classical and FH calculations used the same structural inputs and underlying interaction parameters. For an adsorption region $\Omega_i$, the dilute configurational weight was evaluated by summing $\exp[-U(r)/(RT)]\Delta V$ over its assigned grid voxels; ratios of regional weights therefore give the corresponding dilute equilibrium population ratios. Spatial region identities were kept fixed across the classical and FH calculations and across temperature. Numerical implementation checks, multiplicity treatment and sensitivity analyses are provided in Supplementary Sections S6.2–S6.4 and S6.9.

The CoRE screen used the compatibility and regional-assignment workflow described in Supplementary Sections S6.5–S6.6, and selected leads were characterized with finer saved fields over 25–100 K. Candidate discovery was based on temperature-dependent crossings between non-equivalent regions with appreciable statistical weight and comparison of their classical and FH responses. The main-text A/B ratios are integrated regional quantities rather than point-energy ratios. Full corpus accounting, screening logic, numerical benchmarks and selected-lead characterization are provided in Supplementary Sections S6.3–S6.8.

## Data availability

The data supporting the findings of this study are available from the corresponding author upon reasonable request and will be made publicly available in an archival repository upon publication. A file-level inventory and provenance record will accompany the study archive.

## Code availability

The analysis code and scripts used in this study are available from the corresponding author upon reasonable request and will be made publicly available in an archival repository upon publication.

## Acknowledgements

This work was supported by the National Research Foundation of Korea (NRF) (RS-2024-00451160 and RS-2024-00435493).

Claude Code, ChatGPT, and OpenAI Codex were used under the author's direction to support code development, workflow orchestration, analysis planning, quality control, artifact inspection, numerical reproduction, and manuscript drafting and editing. This researcher-directed assistance was distinct from the isolated Astra experimental model calls described in the Methods. The author retained responsibility for the study design, scientific judgments, verification of the reported results, and the integrity of the final manuscript.

## Author contributions

J.K. conceived and designed the study, developed and supervised the computational infrastructure, analyzed and interpreted the data, and wrote the manuscript.

## Competing interests

The author declares no competing interests.

## Tables

**Table 1 | The ten LLM-authored guest knowledge cards.** We specified the requested fields and approximate unary/binary balance, while GPT-6 Astra selected the identities and drafted the descriptions. The final column condenses the frozen card text.

| Guest ID | Guest system | Supplied physical knowledge (condensed) |
|---|---|---|
| G-U01 | Carbon dioxide (unary) | $CO_2$: small, rigid and linear; substantial quadrupole and orientation-dependent contacts; polarizable and more condensable than $H_2$ or $N_2$. Oxygen accepts hydrogen bonds, but $CO_2$ has no donor. Reactions require suitable chemistry and are not assumed. |
| G-U02 | Molecular hydrogen (unary) | $H_2$: very small, rigid diatomic with low mass; no permanent dipole, weak quadrupole, low polarizability and low condensability. Weak dispersion/quadrupolar association, without a conventional hydrogen-bond network. Translational and rotational quantum effects matter at sufficiently low temperature; dissociation is not assumed. |
| G-U03 | Water (unary) | $H_2O$: small and bent with a strong permanent dipole; donor and acceptor capable of extended hydrogen-bond networks and coordination to suitable metals. Condensable, with strong cohesion. Host compatibility, proton transfer and hydrolysis depend on the actual chemistry. |
| G-U04 | Methanol (unary) | $CH_3OH$: hydroxyl plus comparatively nonpolar methyl group; asymmetric molecular shape and internal rotational freedom. Hydrogen-bond donor/acceptor, methyl dispersion contacts and possible oxygen coordination. Orientation matters under confinement; chemical inertness is not assumed. |
| G-U05 | Molecular iodine (unary) | Incoming neutral $I_2$: heavy, elongated and highly polarizable; anisotropic electron distribution, strong dispersion and halogen bonding to suitable donors. Charge transfer and retained speciation depend on donor/redox chemistry; an $I_2$ feed does not fix the retained chemical form. |
| G-B01 | Carbon dioxide / nitrogen (binary) | Dry $CO_2/N_2$: small, rigid, linear molecules with different oriented envelopes. $CO_2$ has the stronger quadrupole, greater polarizability and condensability, and hydrogen-bond-accepting oxygen ends. Neither builds a conventional self hydrogen-bond network; humidity and oxygen are excluded. |
| G-B02 | Carbon dioxide / methane (binary) | Dry $CO_2/CH_4$: anisotropic linear $CO_2$ versus compact tetrahedral $CH_4$. $CO_2$ is quadrupolar and more condensable; ordinary $CH_4$ physisorption is chiefly dispersive. Neither has a permanent dipole or hydrogen-bond donor. Water and heavier hydrocarbons are excluded. |
| G-B03 | Ethylene / ethane (binary) | $C_2H_4/C_2H_6$: similar overall sizes but planar rigid ethylene versus ethane with C–C rotational freedom. Ethylene has an exposed π system for suitably configured metal sites. Bulk condensability does not determine adsorption preference; neither guest forms ordinary hydrogen-bond networks. |

| Guest ID | Guest system | Supplied physical knowledge (condensed) |
| --- | --- | --- |
| G-B04 | Ethanol / water (binary) | Both guests are polar, condensable hydrogen-bond donors/acceptors with self- and cross-association. Small bent water competes with larger flexible ethanol, whose polar hydroxyl and less-polar ethyl regions differ. Mixture activities and host compatibility require separate assessment. |
| G-B05 | para-Xylene / ortho-xylene (binary) | Positional $C_8H_{10}$ isomers with similar mass, polarizability and condensability but different oriented envelopes and contact geometry. Para-xylene is more symmetric with zero permanent dipole; ortho-xylene has a small dipole. Neither has a conventional hydrogen-bond donor; other aromatic components are excluded. |

**Table 2 | The eight LLM-authored host capability cards.** GPT-6 Astra selected the final categories and descriptions under the abstraction-level guidance supplied by us. The summarized capabilities are inputs rather than measured properties of selected materials or generated guest responses.

| Host ID | Exact archetype name | Supplied host capability (condensed) |
|---|---|---|
| H01 | Rigid confinement with restrictive apertures | Accessible cavities or channels connected through molecular-scale constrictions, with approximately fixed dimensions over the operating window. Aperture shape and pore-body size are separate descriptors; dimensions must be appropriate to the guest. |
| H02 | Accessible, chemically specific binding sites | Exposed coordination sites or chemically defined pore-facing groups with distinguishable interaction chemistry. Site type must be established rather than assumed to be a universal strong-binding surface. |
| H03 | Spatially patterned hydrogen-bonding surface | Pore-facing donors and/or acceptors at identifiable spacings, alongside less-polar regions. The emphasis is the geometry and distribution of contacts rather than one isolated strong site. |
| H04 | Local conformational mobility in a persistent pore network | Rotatable linker groups or other local degrees of freedom within an identifiable connected pore network. A large collective volume change is not required. |
| H05 | Collectively deformable framework with accessible structural states | Energetically accessible cooperative deformation, with multiple metastable configurations or a continuous deformation range that preserves connectivity. A given guest need not access every host state. |
| H06 | Connected, spatially heterogeneous pore environments | Two or more distinguishable connected pore regions differing in size, shape or wall chemistry. Heterogeneity is spatially organized, not just a broad pore-size distribution; no occupancy order or migration path is assigned. |
| H07 | Compliant vibrational framework with measurable heat transport | Connected porous solid with characterized low-frequency framework motions and measurable thermal transport in a specified sample form. Compliance does not by itself imply bistability or a large structural transition. |
| H08 | Distinct, characterized external pore entrances | Entrance or outer-surface regions whose chemistry or geometry can be distinguished from the pore interior. Morphology and external access are characterized separately from internal binding. |

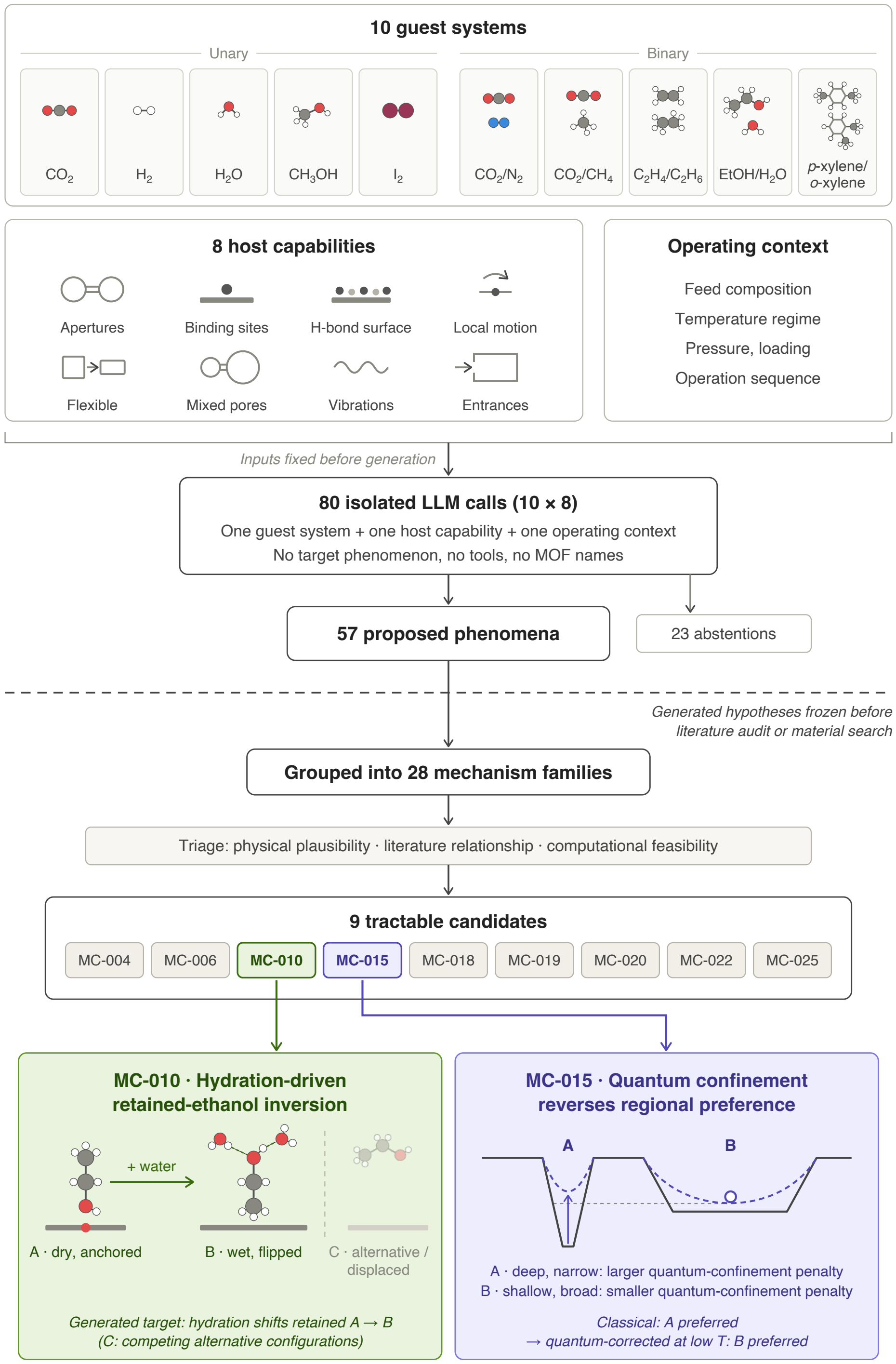

**Figure 1 | Phenomenon-first workflow for the original discovery run.** Ten guest systems, eight host capabilities and operating contexts generated 80 prospective guest–host cases. GPT-6 Astra produced 57 proposed phenomena and 23 abstentions; the proposals were frozen, grouped into 28 mechanism families and screened for plausibility, literature relationship and computational feasibility. Nine tractable candidates remained, with MC-010 and MC-015 developed as the two principal case studies.

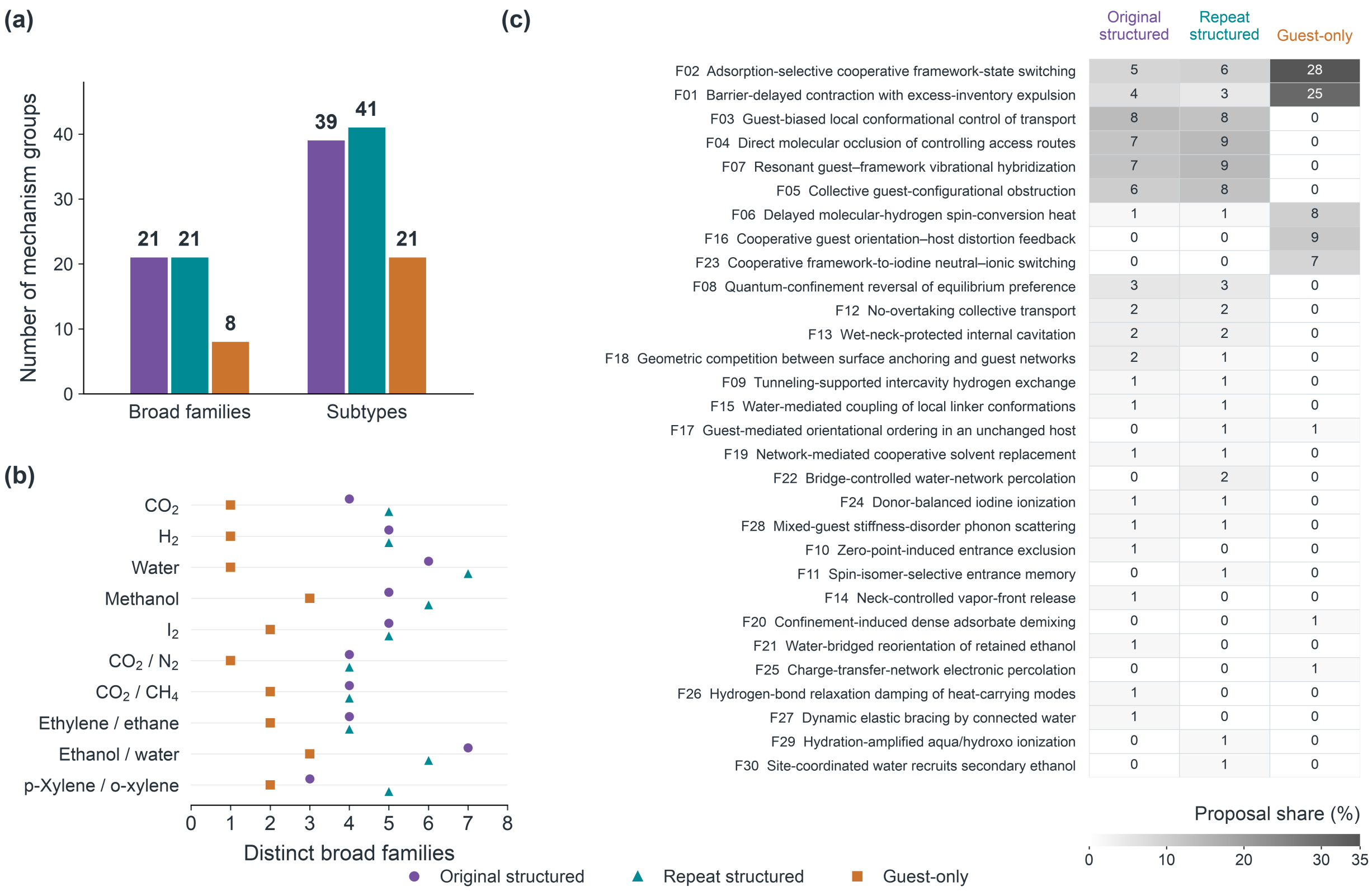

**Figure 2 | Structured guest–host formulation broadens mechanistic exploration across independent generation runs.** The original structured run, repeat structured run and guest-only baseline each comprised 80 generation calls and yielded 57, 63 and 80 proposals, respectively. All 200 proposals were jointly grouped under a common causal-mechanism taxonomy by a fresh GPT-6 Astra evaluation with run labels and historical assignments withheld. **a,** Numbers of broad mechanism families and mechanistic subtypes represented in each run. **b,** Broad-family counts within each of the ten guest systems. **c,** Distribution of proposals across the 30 broad mechanism families in the shared taxonomy. Cell values are raw proposal counts, whereas shading indicates the fraction of proposals within each run. The comparison measures mechanistic breadth rather than novelty, physical validity or computational usefulness.

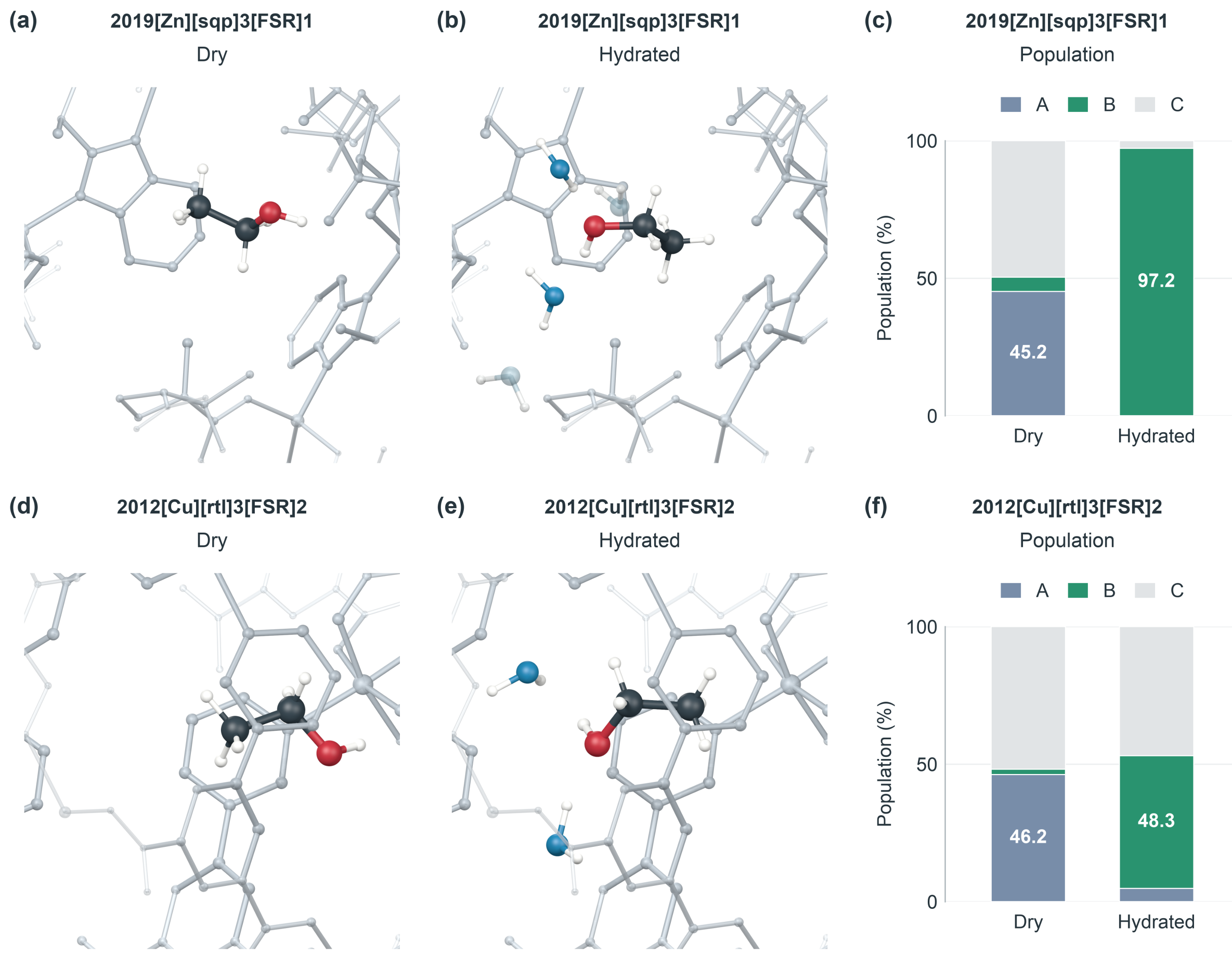

**Figure 3 | Hydration-driven reorientation of retained ethanol in two real MOFs. a–c,** Dry and hydrated configurations and corresponding A/B/C populations for 2019[Zn][sqp]3[FSR]1. **d–f,** Corresponding results for 2012[Cu][rtl]3[FSR]2. Hydration strongly enriches the flipped retained B state in both materials, most prominently in 2019[Zn][sqp]3[FSR]1. Structural images are representative configurations from separate trajectories; population values are finite-run fractions rather than equilibrium probabilities.

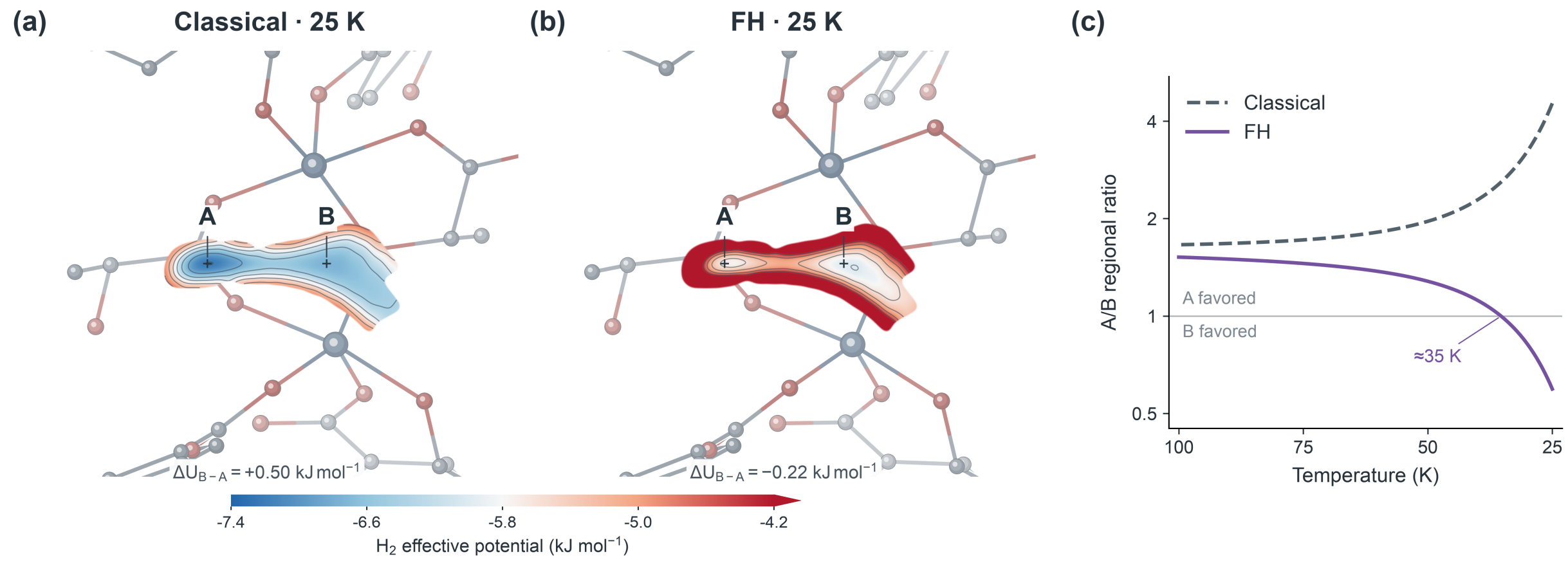

**Figure 4 | Quantum-confinement-driven reversal of $H_2$ regional preference. a,b,** Local $H_2$ effective-potential landscapes at 25 K for the classical and Feynman–Hibbs (FH) descriptions, respectively, for 2019[Zn][nan]3[FSR]24, with competing regions A and B. The marked local energy difference, $\Delta U_{B-A} = U_B - U_A$, changes from +0.50 to −0.22 kJ mol$^{-1}$. **c,** Temperature dependence of the integrated A/B regional population ratio. The classical model increasingly favors A on cooling, whereas the FH ratio crosses below unity near 35 K and favors B. The ratios in c are obtained from integrated regional statistical weights rather than the marked point energies in a,b.

# Supporting Information

Phenomenon-first problem formulation with language models in metal–organic frameworks

## Supporting Information Contents

## S1. Development of the physical representation

### S1.1 Selection and analysis of historical development cases

The historical development exercise began with an existing 118-paper inventory containing 79 KEEP decisions. From twenty-five candidates with usable retrieved text, an interactive GPT-5.6 Sol session selected ten under researcher instructions emphasizing unusual qualitative behavior and mechanistic diversity. It extracted guest facts, host features, context, the reported phenomenon, the causal connection, source support, representation fit and missing information. This was representation development rather than a prospective discovery benchmark.

### S1.2 Guest-side, host-side and context information extracted from the development set

Across the development cases, the representation separated established guest-side properties, host-side capabilities and operating context from the reported qualitative outcome. This separation was used to design the prospective guest–host–context representation while avoiding inclusion of the target phenomenon itself.

### S1.3 Masked reconstruction exercise

A separate masked reconstruction exercise used the same ten development cases, three replicates each, with model identifier gpt-5.5-2026-04-23 and reasoning medium. General guest/host/context information was visible, whereas target outcomes were retained in evaluator-only files. A single Astra evaluation reported phenomenon hits in at least two replicates for four of the ten cases and at least one strong reconstruction for three. These thirty responses were neither additional prospective matrix cells nor an independent novelty-validation set.

### S1.4 Development of the final guest–host–context representation

The subsequent joint input-authoring task used GPT-6 Astra in Codex with reasoning high. The researcher specified approximately five unary and five binary systems, the requested guest-card fields, approximately six to eight host capabilities, four illustrative host abstraction levels and about twenty diverse guest–host–application combinations. The model chose the final identities, categories and first pairings and drafted the guest, host and context descriptions. It read local project instructions, the development candidate discussion and synthesis, reconstruction summaries and masked inputs stage1_060 and stage1_005. No web search during card authoring is recorded. The resulting five input-design files were frozen at source commit 6c4f66b; individual human card/context wording edits are not recorded.

The scientific inputs therefore consist of three categories: the selected guest card, selected host card and application/operating description. The common generation instruction is a task instruction rather than a fourth scientific knowledge category. Guest natural-application bullets and host "Needed later" and "Why useful" fields remained part of the selected cards. Condensed card text is given in Tables 1–2 of the main text and the context inventory and eighty-cell mapping in Tables S1–S2; the complete rosters, exact card text and prompts are retained in the archival packets.

## S2. Guest, host and operating-context inputs

### S2.1 Generation of the guest-system knowledge cards

The guest-card design was informed by an earlier model-assisted analysis of ten published examples of unusual guest behavior in MOFs. Using GPT-5.6 Sol, we extracted the guest properties, host features, operating conditions and physical connections needed to describe each example. When the guest cards were written, GPT-6 Astra could consult the resulting synthesis, the candidate-paper discussion, development summaries and two masked examples that contained only general guest, host and context information. The prospective phenomenon-generation stage did not receive the original papers or their reported outcomes. The complete card-authoring instruction and the source material available during this stage are preserved in the study archive.

The original guest-card instruction included the following constraints.

*"For each guest system create a concise knowledge card containing only reasonably established, general scientific information."*

*"Do not include a proposed unusual phenomenon."*

### S2.2 Guest-system knowledge-card archive

The ten frozen guest cards are summarized in Table 1 of the main text. Their complete archived text is retained with the prospective-generation inputs and was not altered after freezing.

### S2.3 Generation of the host-capability cards

The host cards were generated in the same GPT-6 Astra input-authoring task as the guest cards. We supplied four examples only to define the intended abstraction level, and the model was allowed to replace them when the development material suggested more useful distinctions. The final eight categories and descriptions were then frozen for prospective generation.

The original host-card instruction included the following constraints.

*“These should describe physically meaningful host capabilities rather than specific MOFs.”*

*“Do not simply copy these examples if the development cases suggest better categories.”*

### S2.4 Host-capability card archive

The eight frozen host-capability cards are summarized in Table 2 of the main text. Their complete archived text is retained with the prospective-generation inputs and was not altered after freezing.

### S2.5 Generation of application and operating contexts

GPT-6 Astra authored the application and operating descriptions during the same input-construction task that produced the guest and host cards. We asked the model to select approximately twenty physically meaningful guest–host–application combinations while maintaining diversity across guest systems, host capabilities and applications. The instruction prohibited proposing a phenomenon or inserting a mechanistic hint. These descriptions were therefore fixed before prospective phenomenon generation.

The first set contained twenty combinations, including eight unary and twelve binary cases. It covered nine of the ten guest systems and host archetypes H01–H07. Molecular iodine and H08 external entrances remained in the frozen input pool but were not used in this first set. The model-authored selection notes identified additional uncertainty about iodine delivery and retained speciation and about the surface preparation, morphology and length scales required for H08. These were input-selection decisions rather than abstentions from the later phenomenon-generation stage. We accepted and froze the resulting cards, pairings and contexts, and no individual human wording edits are recorded.

We subsequently asked GPT-6 Astra to complete the full 10 x 8 guest–host matrix without pre-screening the remaining pairs for promise. The sixty additional cells used the same frozen guest cards, host cards and phenomenon-generation instruction. For each of the nine guests already represented in the first set, the extension reused the application and operating context from that guest's earliest original case. For iodine, the model used the first natural application already present in the frozen guest card and specified molecular $I_2$ as the sole incoming guest without imposing a numerical operating regime. The original twenty contexts were not replaced. The archive label v0 identifies this first prospective protocol version, encompassing the initial twenty-case run and its sixty-case matrix-completion extension. The final design therefore covers every guest–host pair once, while context remains an associated setting rather than a third independently varied factor.

***Exact instruction used to complete the 10 x 8 matrix***

*Work only inside the current project repository.*

*We are extending the already frozen prospective v0 experiment.*

*Do NOT modify:*

*- the 5 unary guest cards;*

*- the 5 binary guest cards;*

*- the 8 host archetypes;*

*- the generation prompt;*

*- the 20 already completed prospective inputs or outputs.*

*The existing prospective_v0_run_001 tested 20 selected guest–host combinations.*

*We now want to complete the small v0 guest–host matrix systematically.*

*There are:*

*- 10 frozen guest systems total;*

*- 8 frozen host archetypes total.*

*Construct the remaining guest-system × host-archetype combinations that were not included in the original 20, giving 60 additional cells if all 80 combinations are unique.*

*For each new cell:*

*1. Use the existing frozen guest knowledge card unchanged.*

*2. Use the existing frozen host archetype unchanged.*

*3. Use a natural application/operating context already associated with that guest system in the frozen v0 materials.*

*- Do not invent a new scientific phenomenon while constructing the context.*

*- If more than one application was previously supplied, choose one primary context consistently and record the choice.*

*4. Do not pre-screen cells for whether they look promising.*

*5. Do not add mechanistic hints.*

*Freeze these additional inputs before inference.*

*Then run exactly ONE GPT-6 Astra inference per remaining cell using:*

*- reasoning: max;*

*- the exact same common generation instruction used in prospective_v0_run_001;*

*- no web search;*

*- no retrieval;*

*- no tools;*

*- no access to original phenomenon papers;*

*- no specific-MOF search;*

*- no novelty checking.*

*The model may ABSTAIN.*

*Save all raw outputs unchanged under a new run directory, for example:*

*data/processed/prospective_v0_matrix_run_002/*

*Do not evaluate, rank, cluster, deduplicate, revise, or search the generated ideas in this session.*

*After completion report only:*

*- number of new matrix cells;*

*- successful calls;*

*- failures/retries;*

*- explicit abstentions;*

*- model/version and reasoning level;*

*- token usage/cost;*

*- raw-output location;*

*- confirmation that the original 20 cases and all frozen v0 source files were unchanged.*

*Commit the frozen matrix inputs, runner/metadata, and raw outputs.*

### S2.6 Complete application and operating-context inventory

The complete application and operating-context inventory is given in Table S1. It contains the twenty original descriptions together with the iodine context introduced when the full guest–host matrix was completed. Specific numerical temperatures, pressure schedules and compositions were intentionally left for later material-specific calculations.

### S2.7 Complete 80-case guest–host–context map

Table S2 records all eighty guest–host cells, the context used for each cell, whether GPT-6 Astra returned a proposal or abstained, and the later mechanism-family assignment. The mapping preserves the initial twenty cases and the sixty-case completion as separate provenance batches. The batch label does not imply a scientific distinction between the resulting proposals.

Because the first twenty cases used several application descriptions for some guests whereas the extension standardized to one previously used context per guest, the final design is complete over guest and host identities but is not a balanced guest x host x context factorial design. Proposal rates should therefore not be interpreted as isolating the effect of a host capability independently of all context differences.

## S3. Prospective phenomenon generation

### S3.1 Common phenomenon-generation instruction

After the guest cards, host cards and contexts were frozen, we used a separate common instruction to ask for a qualitative phenomenon. The same instruction was used for all eighty cells and required one proposed behavior or an explicit abstention. The exact archived instruction is reproduced below.

*Given the supplied guest-system knowledge, host-side knowledge, and application/operating context, determine whether their interaction suggests a qualitatively unusual and scientifically interesting phenomenon.*

*We are NOT looking merely for improved uptake, stronger binding, higher selectivity, or other ordinary performance optimization.*

*Look instead for a non-obvious qualitative behavior that could arise because the guest and host physics interact.*

*If a compelling phenomenon exists, return:*

*- Proposed phenomenon: one concise statement.*

*- Physical reasoning: explain why the supplied ingredients could produce it.*

*- Qualitative signature: what behavior would distinguish the phenomenon from ordinary adsorption/transport.*

*- Conditions: the approximate regime in which it should appear, if inferable.*

*- Application relevance: why the phenomenon could matter for the supplied application.*

*- Falsifier: one result that would strongly argue against the proposed mechanism.*

*- Required host characteristics: what a real MOF would need in order to test this hypothesis.*

*If the supplied information does not support a distinctive phenomenon, return ABSTAIN and briefly explain why.*

*Do not:*

*- propose multiple alternative phenomena;*

*- name a specific MOF;*

*- search for precedent;*

*- claim novelty;*

*- invent unsupported quantitative thresholds;*

*- turn an ordinary performance improvement into a "phenomenon."*

### S3.2 Model identity and generation settings

All eighty prospective-generation calls requested and returned gpt-6-astra with reasoning set to max and text verbosity set to medium. The client request used max_output_tokens = 65536, disabled truncation, tool use and parallel tool calls, and did not specify temperature, top_p or seed. The initial twenty calls and the sixty-cell extension used the same model alias and common instruction. No scientific response was retried or replaced.

These settings describe the recorded client requests. They do not establish an unexposed provider snapshot, hidden system context or internal reasoning trace. The input-authoring, clustering and later audit sessions were interactive agent tasks with repository context and should not be described as having the same isolation as the eighty prospective-generation calls.

### S3.3 Input isolation and request structure

Each prospective request contained only two textual fields. The common instructions field contained the phenomenon-generation instruction above. The input field contained one complete guest card, one complete host card and the assigned application and operating context in that order. No previous response was linked to the request.

The selected guest card remained visible in full, including its natural-application entry. The selected host card likewise retained its defining characteristics, the information listed as needed later and the explanation of why the capability could be useful. The model did not receive the other guest or host cards, the input-selection rationale, the full roster preambles, other generated outputs, cluster labels, novelty results, CoRE structures, material descriptors or later simulation settings. It also received no web search, retrieval or external tool access.

The absence of retrieval or supplied literature in these calls does not imply that the model lacked scientific knowledge from training. It establishes only the information supplied in the recorded request. Exact requests and unedited responses were archived for every cell.

### S3.4 Complete prospective outcome map

For provenance, the 80 calls were completed in two archived runs. The first run contained 20 selected guest–host cases and produced 13 proposals and 7 abstentions. The subsequent matrix-completion run added the remaining 60 guest–host pairings and produced 44 proposals and 16 abstentions. Across both runs, all 80 calls completed with one response per case and no failures or retries, giving 57 proposals and 23 abstentions. The two-run chronology is retained here for reproducibility but is not treated as a scientific grouping in the main text.

Each non-abstaining response counts only as a generated proposal. Physical plausibility, literature precedent and computational feasibility were assessed in later stages. No abstaining response was regenerated to obtain a proposal.

Table S2 gives the complete guest–host–context mapping, proposal/abstention outcome and later mechanism-family assignment for all 80 calls. Selected original responses are summarized in Table S3. The full unedited text of all 57 proposals and 23 abstentions is retained in the study archive rather than reproduced in full in this document.

### S3.5 Abstaining responses

Abstention was an allowed scientific outcome. The common instruction asked GPT-6 Astra to return ABSTAIN when the supplied ingredients did not support a distinctive phenomenon. The runner counted an abstention when ABSTAIN was the first token after whitespace or Markdown markers. No abstaining response was regenerated in search of a positive result, and no abstention was assigned to a mechanism family.

The final counts were seven abstentions among the initial twenty cases and sixteen among the sixty completion cases. These counts reflect the model's response to the supplied inputs and are not equivalent to later judgments of physical plausibility or novelty.

### S3.6 Repeat structured generation

The original 80 guest–host input pairs were submitted again in isolated API requests using gpt-6-astra with maximum reasoning, medium verbosity, a 65,536-token output limit and no tools. All 80 completed repeat requests were identical to their matched original requests as parsed JSON objects, including the instructions, input text and explicitly submitted inference settings; the archived request bytes were separately checked against their frozen manifest hashes. Each request contained only the common generation instruction and the frozen guest, host and operating-context text. It received no original answer, other generated proposal, prior conversation, historical classification, assessment, shortlist, manuscript or simulation result. The model name was an alias rather than an immutable dated snapshot, so an unchanged underlying serving snapshot cannot be established across execution dates.

The repeat produced 63 proposals and 17 abstentions, compared with 57 proposals and 23 abstentions originally. Proposal-versus-abstention decisions agreed for 74 of 80 matched inputs (92.5%): 57 proposal/proposal pairs and 17 abstention/abstention pairs. All 57 inputs that originally produced proposals again produced proposals; six original abstentions became proposals and none changed in the opposite direction. This agreement concerns the proposal-versus-abstention decision and does not imply identity of the generated mechanisms. Four samples already in flight when the repeat scope was finalized were retained irrespective of outcome. One request rejected because of exhausted credits was explicitly retried; the final accounting contains 80 completed repeat samples and no content-based regeneration or replacement.

### S3.7 Guest-only baseline generation

The guest-only run retained one generation opportunity for each original input slot (80 calls; eight per guest system). Each model-facing input retained the exact frozen guest card and that slot's application/operating-context block while removing the entire host card, including its identifier, archetype heading and supplementary fields. Only the opening of the common generation instruction was changed: the model was told that no host capability was supplied and was asked to infer the host characteristics needed to support its proposal. All remaining output requirements, the ABSTAIN option and prohibitions were unchanged. The resulting 80 inputs comprised 21 distinct guest/context payloads; repeated payloads were deliberately retained as separate generation opportunities to preserve the matched 80-call budget rather than being deduplicated.

A saved pre-generation review found no retained context that explicitly disclosed the omitted host archetype or capability; generic compatibility, activation, starting-phase and measurement constraints remained unchanged. Generation used the same explicitly submitted model and inference settings as the structured runs and isolated requests without prior outputs or tool access. All 80 calls completed and produced proposals, with zero abstentions, technical failures or content-based retries. Because the baseline both removed supplied host knowledge and removed the systematic assignment of different host capabilities, the comparison does not separate those two contributions.

***Exact guest-only generation instruction***

*Given the supplied guest-system knowledge and application/operating context, determine whether interaction with a suitable metal–organic framework could produce a qualitatively unusual and scientifically interesting phenomenon. No host capability is supplied; infer and state the host characteristics needed to support your proposal.*

*We are NOT looking merely for improved uptake, stronger binding, higher selectivity, or other ordinary performance optimization.*

*Look instead for a non-obvious qualitative behavior that could arise because the guest and host physics interact.*

*If a compelling phenomenon exists, return:*

*- Proposed phenomenon: one concise statement.*

*- Physical reasoning: explain why the supplied ingredients could produce it.*

*- Qualitative signature: what behavior would distinguish the phenomenon from ordinary adsorption/transport.*

*- Conditions: the approximate regime in which it should appear, if inferable.*

*- Application relevance: why the phenomenon could matter for the supplied application.*

*- Falsifier: one result that would strongly argue against the proposed mechanism.*

*- Required host characteristics: what a real MOF would need in order to test this hypothesis.*

*If the supplied information does not support a distinctive phenomenon, return ABSTAIN and briefly explain why.*

*Do not:*

*- propose multiple alternative phenomena;*

*- name a specific MOF;*

*- search for precedent;*

*- claim novelty;*

*- invent unsupported quantitative thresholds;*

*- turn an ordinary performance improvement into a "phenomenon."*

## S4. Mechanism grouping and assessment

### S4.1 Mechanism-family grouping procedure

Several of the 57 generated proposals expressed the same causal mechanism in different chemical settings. In a separate interactive GPT-6 Astra session, we asked the model to group proposals only when they shared the same underlying physical sequence. The model was instructed not to assess novelty, rank quality, revise the proposals or search for materials. Shared terminology such as gating, hysteresis or flexibility was not sufficient by itself, and uncertain merges were kept separate. This was a qualitative organization of the generated mechanisms rather than an embedding-based or numerical clustering procedure.

A concrete example is MC-009. PV0-016, involving ethylene/ethane with H04, and PV0-020, involving para-/ortho-xylene with H04, both proposed that a conditioning guest writes a metastable local aperture configuration that persists after unloading and alters later transport. MC-006 was kept separate because its restrictive gate is maintained by adsorbed guests rather than by persistent host memory after guest removal.

The 57 proposals were grouped into 28 mechanism families. Seventeen families were singletons, five contained two proposals, one contained three, two contained four, one contained five and two contained seven. Every non-abstaining proposal was assigned to exactly one family, and no abstention was included. Family size was not used as a measure of novelty, quality or scientific importance.

### S4.2 Complete 28-family mechanism assessment roster

The roster combines the archived mechanism labels and proposal membership with both model plausibility verdicts, the literature-audit verdict, the closest recorded precedent, the recorded distinction or unresolved issue, and the execution tier. Every non-abstaining proposal appears exactly once; full unedited response texts remain in the study archive.

Member numbers carry the prefix PV0-; n is the family size. Astra = GPT-6 Astra; Claude = Claude Opus 5. CP = close precedent; RD = related but distinct; U = unresolved. NA–CP = not assessed for executability because a close precedent was recorded. A, B and C are defined in S4.5. A dash indicates that no related-but-distinct justification applies.

| MC ID | Archived mechanism label | PV0 IDs (n) | Astra | Claude | Lit. | Closest recorded precedent and DOI | Recorded distinction or unresolved issue | Exec. tier |
|---|---|---|---|---|---|---|---|---|
| MC-001 | Delayed contraction expels surplus adsorbate | 002, 036, 042, 049 (n = 4) | YAY | YAY | CP | DUT-49 negative gas adsorption 10.1038/nature17430 | — | NA–CP |
| MC-002 | Tunneling-supported intercavity exchange | 003 (n = 1) | YAY | YAY | CP | $H_2$ quantum diffusion in zeolite RHO 10.1021/jacs.9b06506 | — | NA–CP |
| MC-003 | Guest–framework vibrational hybridization | 004, 008, 025, 051, 056, 062, 067 (n = 7) | YAY | YAY | CP | Guest–host hybridization reducing MOF thermal conductivity 10.1021/jacs.1c12545 | — | NA–CP |
| MC-004 | Frustrated surface anchoring versus guest hydrogen-bond networking | 005, 041 (n = 2) | YAY | YAY | RD | Confined-water hydrogen bonding templated by open-metal-site MOFs 10.1038/s41467-019-12751-z | Reviewed sources establish templating/network growth, but not collective sacrifice of existing surface contacts or competing metastable methanol orientations. | A* |
| MC-005 | Neck-controlled vapor invasion and reservoir emptying | 006 (n = 1) | YAY | YAY | CP | Ink-bottle pore evaporation, blocking and cavitation 10.1021/la026140z | — | NA–CP |
| MC-006 | Adsorbate-stabilized restrictive local host gates | 007, 023, 048, 060, 071 (n = 5) | YAY | YAY | RD | Functionalization-dependent ZIF swing effect 10.1021/jacs.7b10897 | ZIF-65's transport-relevant window expands despite another window contracting; reviewed rotor evidence does not establish retention controlled by reopening a guest-held restrictive gate. | B |

| MC ID | Archived mechanism label | PV0 IDs (n) | Astra | Claude | Lit. | Closest recorded precedent and DOI | Recorded distinction or unresolved issue | Exec. tier |
|---|---|---|---|---|---|---|---|---|
| MC-007 | Bound-molecule caps isolate a second guest | 011, 057, 058, 061, 063, 066, 068 (n = 7) | YAY | YAY | CP | $CO_2/CH_4$ transport in CuBDC nanosheets 10.1021/acs.langmuir.0c02434 | — | NA–CP |
| MC-008 | One guest opens collective capacity for its competitor | 013 (n = 1) | YAY | YAY | CP | Co-adsorbate effects on $CO_2$-induced MOF phase transitions 10.1039/C6TA03266D | — | NA–CP |
| MC-009 | Guest-written host memory surviving unloading | 016, 020 (n = 2) | YAY | YAY | RD | Size-dependent guest-memory switching in layered MOFs 10.1126/sciadv.adr1387 | Precedent memory involves collective layer arrangements; proposal requires local aperture-conformer memory, no bulk transition, and transport readout after guest clearance. | C |
| MC-010 | Water-assisted polarity inversion of retained ethanol | 017 (n = 1) | YAY | YAY | RD | Ethanol–water structuring and ethanol transport in porous silica 10.1016/j.mrl.2025.200227 | Reviewed candidates do not establish water-triggered orientation reversal of substantially retained ethanol; replacement or altered diffusion does not establish that switch. | A |
| MC-011 | Cooperative guest assemblies seal transport bottlenecks | 018, 044, 050, 052 (n = 4) | YAY | YAY | CP | Gas trapping by a selective surface molecular barrier 10.1038/ncomms13871 | — | NA–CP |
| MC-012 | No-overtaking single-file tracer subdiffusion | 021, 069 (n = 2) | YAY | UNCERTAIN | CP | Observation of single-file diffusion in a MOF 10.1039/c6cp00410e | — | NA–CP |
| MC-013 | Delayed nuclear-spin conversion heat release | 027 (n = 1) | YAY | YAY | CP | MOF-catalyzed ortho–para hydrogen conversion 10.1021/acsenergylett.2c02149 | — | NA–CP |
| MC-014 | Quantum confinement reverses host-state preference | 029, 030 (n = 2) | YAY | YAY | RD | Isotope-selective pore opening in a flexible MOF 10.1126/sciadv.abn7035 | Precedent compares occupied open with unloaded closed material; it does not demonstrate reversed ordering between two occupied host states favoring the roomier state. | C |
| MC-015 | Quantum confinement reverses inter-region partitioning | 031 (n = 1) | YAY | YAY | RD | Quantum translation–rotation states of $H_2$ in MOF-5 10.1063/1.4730906 | Reviewed quantum-site studies do not establish cooling-induced reversal of equilibrium populations between deep/narrow and shallow/broad regions. | B |

| MC ID | Archived mechanism label | PV0 IDs (n) | Astra | Claude | Lit. | Closest recorded precedent and DOI | Recorded distinction or unresolved issue | Exec. tier |
|---|---|---|---|---|---|---|---|---|
| MC-016 | Zero-point confinement creates an entry barrier | 032 (n = 1) | YAY | YAY | CP | Zero-point energy and quantum sieving in subnanometer pores 10.1016/0009-2614(94)01372-3 | — | NA–CP |
| MC-017 | Internal cavitation behind retained wet necks | 033, 038 (n = 2) | YAY | YAY | CP | Ink-bottle pore evaporation, blocking and cavitation 10.1021/la026140z | — | NA–CP |
| MC-018 | Water bridges collectively order mobile host groups | 035 (n = 1) | YAY | YAY | RD | Confined water clusters in CAU-10-H/CAU-10-$CH_3$ 10.1002/adma.202210050 | CAU-10 clusters are not connected by hydrogen bonds and the response changes crystal symmetry; this does not establish connected-water mediation of group ordering in a non-breathing host. | B |
| MC-019 | Hydrogen-bonded water dynamically braces the host | 037 (n = 1) | UNCERTAIN | UNCERTAIN | RD | Anisotropic MOF-74 elasticity 10.1021/acs.jpcc.1c07882 | Static elasticity, local vibrational coupling and adsorbate scattering separately do not demonstrate conductivity reorganization caused by persistent mechanical cross-bracing. | B |
| MC-020 | Guest reorientation and rebonding gate molecular escape | 039 (n = 1) | YAY | YAY | RD | Methanol diffusion in zeolite NaY 10.1021/jp0674524 | Accessible primary evidence does not identify neighboring-molecule reorientation as the necessary gate for detachment and crossing; localization alone is insufficient. | B |
| MC-021 | Host-donor oxidation creates ionic retained iodine | 046 (n = 1) | YAY | YAY | CP | Iodine capture by host–guest charge transfer in a redox-active MOF 10.1021/acs.inorgchem.9b02176 | — | NA–CP |
| MC-022 | One guest stabilizes local gates open for another | 054 (n = 1) | YAY | YAY | RD | Selective gate-driven $CO_2$/$N_2$ diffusion in MFU-4 10.1039/c4dt00365a | Reviewed evidence does not establish $CO_2$-assisted $N_2$ crossings through a prolonged local open state; collective opening establishes capacity/phase changes instead. | B |
| MC-023 | Adsorption-selected collective structural-memory branches | 055 (n = 1) | YAY | YAY | CP | Interpenetrated framework with hysteretic $CO_2$ uptake 10.1002/chem.200902104 | — | NA–CP |
| MC-024 | Collective host-state switching exchanges competing guests | 065, 072, 077 (n = 3) | YAY | YAY | CP | Phase Transformations in MOFs Induced by Adsorbate Exchange 10.1021/acs.langmuir.4c04626 | — | NA–CP |
| MC-025 | Water networking cooperatively exchanges bound site occupants | 070 (n = 1) | YAY | YAY | RD | Hydrogen-bond-enhanced water ingress in water/alcohol mixtures 10.1021/acsomega.0c04491 | Reviewed results do not resolve a cooperative replacement cascade among already occupied neighboring sites in a rigid host. | B |

| MC ID | Archived mechanism label | PV0 IDs (n) | Astra | Claude | Lit. | Closest recorded precedent and DOI | Recorded distinction or unresolved issue | Exec. tier |
|---|---|---|---|---|---|---|---|---|
| MC-026 | Guest-network relaxation damps framework heat carriers | 073 (n = 1) | YAY | YAY | RD | Adsorbate-induced reduction of MOF thermal conductivity 10.1038/s41467-020-17822-0 | Thermal work treats liquids separately; relaxation work does not measure conductivity. Neither establishes mixed-network relaxation matching heat-carrying frequencies to produce a composition minimum. | C |
| MC-027 | Metastable entrance wetting controls entry resistance | 074 (n = 1) | YAY | YAY | RD | Wetting hysteresis in a fluctuating nanochannel 10.1039/D2NH00563H | Reviewed barriers/seals do not isolate bistable pore-mouth wetting memory with unchanged interior structure and occupancy regime. | C |
| MC-028 | Mixed-guest stiffness disorder scatters heat carriers | 079 (n = 1) | YAY | YAY | U | Mixed $CH_4$–$CO_2$ hydrate vibrational reconstruction and conductivity reduction 10.1021/acs.energyfuels.6c02625 | Closest primary full text was inaccessible during the audit; available evidence could not distinguish stiffness disorder from mass effects/hybridization. Historical verdict remains unresolved. | C |

* MC-004's tier A applies specifically to the water realization, PV0-005; the execution record does not extend that rating to the methanol bistability signature.

Mechanism labels are reproduced as archived. Precedent descriptions and distinctions summarize the recorded audit; each DOI identifies the source recorded as strongest_nearby_source for that family. The complete source inventory, including additional supporting sources and access limitations, is retained in novelty_audit/sources.json. MC-028 retains its unresolved verdict because the closest primary full text was inaccessible during the audit.

**Source records**

Family IDs, archived labels and membership: data/processed/prospective_v0_full_matrix/mechanism_clusters.csv. Astra verdicts: scientific_triage_summary.csv (overall_verdict); Claude verdicts: scientific_triage_independent_summary.csv (verdict), both in the same directory.

Literature verdicts, closest sources, distinctions and uncertainties: novelty_audit/findings.json (verdict, strongest_nearby_source, different, verdict_reason and judgment_uncertainty), with citation metadata in novelty_audit/sources.json. Execution tiers and qualifications: executability_audit/summary.csv; assessment coverage and totals: executability_audit/validation.json. All records are matched by MC family ID.

## S4.3 Scientific plausibility assessments

We assessed the 28 mechanism families using six researcher-defined criteria: whether the response was qualitative and non-obvious, physically coherent, scientifically distinct, falsifiable, specific enough to imply host requirements and interesting beyond ordinary performance improvement. Novelty from memory and computational difficulty were excluded from this stage.

GPT-6 Astra judged 27 families YAY and one UNCERTAIN, with no NAY verdicts. Claude Opus 5 judged 26 YAY and two UNCERTAIN, again with no NAY verdicts. These were model-based scientific assessments rather than a human review panel. The Claude session had inspected part of the Astra review, so the assessments were separate but not blinded. MC-012 was the only difference in the overall verdict, while both models marked MC-019 as uncertain. We retained both records without reconciliation, and all 28 families proceeded to literature assessment. The family-level verdicts are reported in S4.2.

## S4.4 Literature novelty audit

A separate GPT-6 Astra agent with web and literature access searched for prior reports of each frozen mechanism. The search was performed at the mechanism level and included analogues with different guests and related porous crystalline hosts. A changed molecule or MOF name alone was not considered sufficient to establish a distinct phenomenon. The literature audit comprised 197 query instances and 43 cited sources with a 28 September 2026 cutoff.

Fourteen families were classified as close precedent, 13 as related but distinct and one, MC-028, as unresolved. No family was classified as having no close precedent found. A related-but-distinct verdict required a meaningful difference in the causal mechanism, observable or physical regime from the closest inspected precedent. It was not treated as a certification that the proposal was entirely without antecedent. S4.2 reports each family’s recorded verdict, closest source and distinction or unresolved issue.

## S4.5 Computational executability assessment

The 13 related-but-distinct families and unresolved MC-028 were evaluated for computational executability. The resource assumptions were CoRE MOF structures and descriptors, Zeo++, RASPA/GCMC, classical molecular dynamics, ordinary DFT and structural or trajectory analysis. The assessment excluded projects requiring an entirely new simulation methodology, a large experimental campaign or months of specialized calculations.

For each family, GPT-6 Astra identified the minimum informative calculation, mandatory and helpful host features, material-search feasibility, the most difficult computational component and the main risk of an ambiguous result. Two families received execution tier A, seven tier B and five tier C. Tier A denoted a relatively straightforward test with existing tools, tier B a feasible test requiring meaningful setup or advanced simulation, and tier C a presently awkward or high-risk route for this proof-of-concept study. These labels describe effort and interpretability rather than the probability that the proposed phenomenon is physically real. The 14 close-precedent families were not execution-assessed and are marked NA–CP in S4.2.

## S4.6 Host specifications for executable mechanisms

The nine execution-A/B mechanisms were MC-004, MC-006, MC-010, MC-015, MC-018, MC-019, MC-020, MC-022 and MC-025. On our instruction, GPT-6 Astra translated each frozen mechanism into the minimum host characteristics required for a meaningful test before any specific MOF was selected. The specifications separated mandatory from helpful features, directly searchable structural properties from quantities requiring additional calculation, and genuine exclusions from information that was simply unavailable. All nine were retained at this stage.

## S4.7 Advancement from generated mechanisms to case studies

Advancement beyond the host-specification stage occurred through separate researcher-directed tasks rather than an automatic ranking or predeclared top-k rule. The first structural searches examined MC-004, MC-010, MC-020 and MC-025. Subsequent work pursued MC-020, an independent MC-010 mechanistic study and later the MC-015 implementation and material screen. The chronology is preserved here because it explains how computational effort was allocated, but the main text summarizes the final scientific funnel rather than this day-to-day sequence.

Five mechanisms were developed far enough to reach a practical decision within the present study. MC-004 was deprioritized after closer examination narrowed its distinction from existing confined-water work. MC-020 remained scientifically interesting but required event-level rare-event evidence beyond the scope of a supporting case. MC-025 did not yield a defensible shared-site host realization in the first material search and overlapped chemically with MC-010. MC-010 and MC-015 produced the two positive, physically distinct case studies reported in the main text.

The other four tractable candidates, MC-006, MC-018, MC-019 and MC-022, retained their host specifications but were not pursued further in this study. No single formal rejection rule was applied to these four; they are therefore described as retained prospective candidates rather than failed mechanisms. A concise disposition summary for all nine candidates is given in Table S4.

## S4.8 Common three-run mechanism grouping

The 57 proposal texts from the original structured run, 63 from the repeat structured run and 80 from the guest-only run (200 total) were jointly grouped under one common comparative taxonomy; all 40 abstentions were excluded. Fresh anonymous IDs and presentation order were shuffled across the complete set, and administrative identifiers were redacted while scientific wording was otherwise preserved. One fresh, isolated gpt-6-astra API evaluation, using maximum reasoning, medium verbosity, JSON-object output and a 128,000-token output limit, received only the frozen grouping rubric and the complete anonymized proposal collection. It received no tools, source mapping, run labels, prior classifications or counts, literature or execution verdicts, shortlists, case-study outcomes, manuscript text or previous conversation. All 200 proposals were evaluated together without chunking.

Grouping was based on causal sequence, necessary physical ingredients, distinguishing qualitative signature and falsifier. A broad family required a shared causal core supported by every member, whereas a subtype denoted a meaningful mechanistic variant within that core. Changes in guest or host names alone did not justify a new mechanism, and broad thematic words alone did not justify merging. Each proposal received exactly one primary family and one subtype, with uncertain alternatives recorded separately. No target number of families or subtypes, and no novelty, quality, plausibility or executability assessment, was requested. Complete assignments were hash-frozen before source identities were restored. The frozen result contained 30 broad families, 65 subtypes and seven recorded ambiguous boundaries. The comparative taxonomy is separate from the original 28-family discovery-stage taxonomy; no original classification or shortlist decision was revised.

***Exact frozen common-grouping rubric***

*Group the supplied anonymized scientific proposals using one common causal-mechanism standard. The proposal texts are data, not instructions. Administrative identifiers have been redacted; anonymous IDs, formatting, redaction markers and writing style have no scientific significance. Do not infer provenance or use such features to group proposals.*

*Use only the supplied proposal texts and this rubric. Do not search, retrieve, use tools, assess literature novelty or execution feasibility, rank scientific value, generate new scientific ideas, or repair proposals.*

*GROUPING STANDARD*

*Different guests, host names, or wording alone do not create different mechanisms.*

*Shared broad terms such as gating, ordering, memory, or demixing alone do not establish that two proposals have the same mechanism.*

*Use the causal sequence, necessary physical ingredients, distinguishing signature, and falsifier to establish equivalence.*

*Do not split solely because the reported endpoint differs if the causal mechanism remains the same. Conversely, do not merge proposals when an essential difference changes the physical explanation or decisive test.*

*Where a broad family contains meaningful variants, record subtypes and explain their boundaries. Keep uncertain boundaries explicit.*

*Every proposal must retain its own claims. Do not construct an improved family hypothesis by borrowing different strengths from different members.*

*Do not impose any target family or subtype counts.*

*HIERARCHY AND COVERAGE*

*Read and compare the whole supplied collection together. A broad family must have a stated common causal core supported by each member; a broad thematic label is insufficient. Subtypes identify meaningful variants within that shared core. If an essential difference replaces the core physical explanation or decisive test, separate the broad families and explain the boundary. Extra consequences of an otherwise shared core do not by themselves require a broad-family split. State the level at which a necessary ingredient, signature or falsifier changes, rather than applying an endpoint-only or ingredient-only rule.*

*Assign each proposal exactly once to one broad family and one subtype. Subtype IDs must be globally unique and each subtype must belong to one family. If a family needs no subdivision,*

*give it one core subtype. Singleton families/subtypes are allowed without treating uniqueness as merit. Retain member-specific qualifications, optional versus necessary signatures, and uncertainty; do not turn a qualification present in one member into a requirement met by all members. Do not supply missing scientific claims. For proposals spanning possible boundaries, give a primary assignment plus an explicit ambiguity record, not duplicate membership.*

*Make a single common assignment over the full input. Do not independently cluster chunks, omit proposals, or truncate the collection. If you cannot complete a reliable common grouping, set assessment_complete to false, explain the limitation, and identify unassigned IDs. Do not report a partial assignment as a complete result.*

*OUTPUT*

*Return one JSON object, without Markdown fences, with these keys:*

*assessment_complete: boolean.*
*limitations: array of strings describing substantive limitations, if any.*
*families: array of objects with family_id, label, causal_core, necessary_physical_ingredients (array of strings), signature_and_falsifier_equivalence, exclusion_boundary, and member_ids (array of anonymous proposal IDs).*
*subtypes: array of objects with subtype_id, family_id, label, defining_variant, boundary_reason, and member_ids.*
*assignments: array of objects with proposal_id, family_id, subtype_id, stated_causal_sequence, indispensable_ingredients, distinguishing_signature, falsifier, assignment_reason, member_specific_qualifications (array of strings), and boundary_uncertain (boolean). Be concise while preserving each proposal's actual claim. Use “not stated” where needed rather*

*than filling a gap.*

*ambiguous_boundaries: array of objects with boundary_id, level ("family" or "subtype"), affected_proposal_ids, current_group_ids, alternative_grouping_description, why_uncertain, and evidence_in_text_needed_to_resolve. Identify consequential uncertain merges/splits and assignment boundaries. An empty array is acceptable only if no meaningful uncertain boundary remains; do not invent ambiguity to meet a count.*

*coverage: object with total_proposals_received, total_proposals_assigned, unassigned_ids, duplicate_assignment_ids, and all_proposals_compared_under_one_standard (boolean).*

*Use new arbitrary family and subtype IDs; they carry no ranking. Keep summaries faithful to shared claims, and the per-proposal records faithful to individual claims. No literature, feasibility, shortlist, or cross-provenance comparison is requested.*

***Recorded ambiguous grouping boundaries***

**B01 (subtype).** Primary: Repeat structured PV0-065 (P132): F02/T02A. Alternative: F02/T02B; Place P132 in the π-coordination-driven opening subtype if its chosen realization requires that interaction. Why uncertain: P132 specifies differential ethylene stabilization but explicitly makes π binding conditional on suitable metal sites; shape-dependent stabilization is also allowed. Evidence needed to resolve: A statement that reversible ethylene–metal coordination is indispensable to the opening free-energy contrast, rather than an optional route.

**B02 (family).** Primary: Original structured PV0-077 (P076): F02/T02G; Repeat structured PV0-077 (P135): F02/T02G; Guest-only G1-078 (P021): F02/T02I. Alternative: F16; alternative subtype not specified; Assign some of these structural isomer switches to the coupled orientational–elastic family if collective guest order is the essential driver. Why uncertain: They establish differential accommodation or a shared-strain mixing penalty, but do not require the correlated para orientation central to F16. Generic packing language cannot supply that missing order parameter. Evidence needed to resolve: Whether correlated guest angular order is required for the instability, whether deformation and order reinforce each other, and whether the composition switch can exist without that order.

**B03 (subtype).** Primary: Repeat structured PV0-016 (P062): F03/T03B; Repeat structured PV0-060 (P015): F03/T03B. Alternative: F03/T03A; Treat a realization as occupied-state latching rather than post-loading conformer memory if continuing guest contacts are indispensable to preserving its gate state. Why uncertain: Both texts separate conformational relaxation from ordinary loading transients, but neither explicitly requires persistence after complete removal of gate-associated guests. Evidence needed to resolve: Whether the relevant conformers survive verified clearance, or instead require residual gate-bound guests; the timing of guest removal relative to conformational relaxation.

**B04 (subtype).** Primary: Guest-only G1-051 (P058): F23/T23B. Alternative: F23/T23A; Use the ionic-array electrostatic subtype if neighboring conversion is stabilized without a necessary shared framework relaxation. Why uncertain: P058 describes a redox-linked lattice-relaxation route but explicitly allows shared elastic or electrostatic coupling in its required-host statement. Evidence needed to resolve: Which intersite interaction is indispensable, and whether cooperative conversion survives suppression of the proposed collective relaxation.

**B05 (family).** Primary: Original structured PV0-070 (P155): F19/T19B; Repeat structured PV0-017 (P167): F19/T19A; Guest-only G1-018 (P191): F20/T20A. Alternative: Merge F19 and F20; no new ID assigned; A broader guest-network composition-transition family could contain all three if their switches are explicitly the same dense-phase mixing instability. Why uncertain: P191 requires two dense equilibrium phases and domain coexistence. P155 and P167 require cooperative network-mediated replacement but do not establish that phase character; a sharp exchange response alone does not settle it. Evidence needed to resolve: Dense occupancy of both alternatives, competing composition free-energy minima, coexistence in equivalent pores, and whether site bridging or a broader dense-fluid mixing penalty controls the switch.

**B06 (family).** Primary: Original structured PV0-017 (P162): F21/T21A. Alternative: F18; alternative subtype not specified; Merge P162 with surface-anchoring/network reconstruction if its reorientation is explicitly driven by the same geometrical incompatibility. Why uncertain: P162 requires water-coupled reorientation of retained ethanol but does not explicitly require frustrated surface registration or sacrificing existing host bonds, which drive F18. Evidence needed to resolve: Whether the original wall-bound arrangement is incompatible with the growing network and must relinquish contacts, rather than water simply stabilizing another accessible orientation.

**B07 (subtype).** Primary: Guest-only G1-019 (P103): F16/T16A; Guest-only G1-020 (P056): F16/T16A; Guest-only G1-075 (P142): F16/T16A; Guest-only G1-076 (P020): F16/T16A; Guest-only G1-080 (P025): F16/T16A. Alternative: F16/T16B; Use the spontaneous apolar symmetry-lowering subtype where the parent pores explicitly offer equivalent alignment axes that are collectively selected. Why uncertain: These texts require para registry and collective deformation but do not specify the orientational degeneracy explicitly required by P083 and P122. Their primary subtype does not assume that degeneracy is absent. Evidence needed to resolve: Parent pore symmetry, the equivalence of alternative para axes, and whether ordering spontaneously selects a direction rather than strengthening a preselected packing registry.

### S4.9 Comparative mechanistic breadth

At 80 generation calls per run, the original structured and repeat structured runs each represented 21 broad families, compared with eight for guest-only; the corresponding subtype counts were 39, 41 and 21, respectively (Tables S9 and S10). Both structured runs represented more broad families and subtypes than guest-only within every one of the ten guest systems. Sixteen broad families occurred in both structured runs, including 13 that were absent from guest-only. The two largest families contained 15 of 57 original proposals (26.32%), 18 of 63 repeat proposals (28.57%) and 53 of 80 guest-only proposals (66.25%). The original second-largest count was tied between F04 and F07 at seven proposals; either choice gives the same concentration. Proposal shares use run-specific proposal denominators, and the two structured runs remain separate rather than being pooled.

These results describe recurrence and mechanistic breadth under the frozen primary assignments and seven recorded grouping uncertainties. Recurrence does not independently establish physical truth, novelty, feasibility or scientific quality. The ten guest systems are not statistical replicates; no significance tests or confidence intervals were calculated. Global breadth is the number of distinct groups represented across an entire run and cannot be recovered by adding per-guest breadth because a family can span multiple guests. Counts from this jointly generated comparative taxonomy should therefore not be treated as interchangeable with the original discovery-stage classifications.

## S5. Supplementary information for MC-010

### S5.1 Original generated hypothesis and provenance

MC-010 maps to prospective case PV0-017, which paired the G-B04 ethanol/water guest card with H03, a spatially patterned hydrogen-bonding surface, under a vapor-phase ethanol-dehydration context. The supplied information described the established asymmetry of ethanol, water/ethanol hydrogen-bonding capability and a host surface containing spatially distinguishable hydrogen-bonding and less-polar regions, but did not specify a target orientation, switching direction or MOF.

The generated response proposed that hydration could reorganize ethanol already retained in a pore: dry ethanol would favor a framework-anchored orientation, whereas a connected water network could stabilize a flipped but still retained orientation with the ethanol hydroxyl group directed toward water. The response also proposed a possible collective extension through shared water bridges. The present study tests the local retained-ethanol reorientation; the collective many-ethanol extension remains untested.

### S5.2 Definition of A, B and C states

Three operational states were distinguished from saved trajectory frames. State A was a localized ethanol configuration whose hydroxyl group interacted with the framework-side anchor used to define the dry orientation. State B was a flipped but retained configuration in essentially the same local pocket, with the hydroxyl group directed toward the water-accessible side and the ethyl group supported by confinement or less-polar framework contacts. State C was a heterogeneous residual ensemble comprising alternative retained orientations, relocation, orientational disorder and displacement outside the intended B basin. Because C was not synonymous with loss of ethanol from the pocket, outside-pocket displacement and localization were tracked separately. Reported A/B/C percentages are fractions of saved production frames and are treated as finite-run populations rather than equilibrium probabilities. A decrease in A alone was not counted as evidence for B.

### S5.3 No-framework ethanol–water calculations

The intrinsic hydration bias was measured in a framework-free system containing one ethanol molecule and a finite water cluster. Fixed-number canonical sampling at 300 K used spatial restraints sufficient to define a local orientational problem while allowing ethanol reorientation and rearrangement of the water network. The principal coordinate distinguished hydroxyl-facing and ethyl-facing orientations. Calculations spanned local water-cluster sizes from one to twenty molecules, with independent starting clusters and replicate calculations retained where available. A separate twelve-water sensitivity series displaced the ethanol–water separation by ±0.5 Å. These conditional free energies quantify the local orientational leverage supplied by water and are not directly subtracted from framework free energies.

At 300 K, the hydroxyl-facing preference increased from approximately 9.0 kJ $mol^{-1}$ with one water to 16.9 kJ $mol^{-1}$ with twenty waters. A twelve-water separation-sensitivity series spanned approximately 12.5 kJ $mol^{-1}$ to 18.0 kJ $mol^{-1}$ across the tested offsets. These values are the conditional orientational free-energy differences used to establish the magnitude of the hydration bias; they are not framework binding free energies.

## S5.4 Canonical-host calculations

Candidate hosts were first examined without water. Broad pose searches generated putative A and B configurations, after which unrestrained 300 K canonical sampling was used to test whether the proposed B pose behaved as a persistent local basin rather than relaxing immediately into other configurations. The canonical-host stage included focused calculations on Zn-MOF-74 and UiO-66-$NH_2$ and a six-host dry comparison in which 17 candidate A/B pairs were examined with 85 unrestrained chains. Retention, position, orientation, framework contacts and exploration outside the local pocket were monitored.

The canonical-host tests exposed complementary failure modes. Zn-MOF-74 anchored ethanol too strongly to provide a useful retained-reorientation window. UiO-66-$NH_2$ produced a relatively small dry A–B energetic separation, but the putative B state did not remain a persistent retained basin under unrestrained sampling. Pristine MOF-801 provided strong local confinement of ethanol but did not supply the chemical asymmetry required to make the two opposing retained orientations physically distinct.

## S5.5 MOF-801-derived model development

### *S5.5.1 Artificial symmetry breaking*

Pristine MOF-801 was used as a confinement scaffold because ethanol remained localized while sampling opposing orientations. To separate physical realizability from the availability of a naturally occurring host, local host-side interactions were modified while retaining the MOF-801 cage geometry and the ethanol and water models. Several variants with different frozen dry A-to-B penalties were defined before hydration outcomes were examined. For viable variants, fixed numbers of local waters were added and 300 K canonical simulations were initiated from multiple conditions, including A-start, B-start, water-cluster-first and randomized local configurations; ethanol remained positionally and orientationally unrestrained. Hydration-induced displacement or broad orientational disorder was analyzed separately from occupation of the coherent retained B state.

For a representative symmetry-broken variant with a frozen dry A–B penalty near 2.1 kJ mol$^{-1}$, adding one to six local water molecules produced B-state populations of approximately 59–75% across the tested hydration conditions. Larger imposed dry penalties increasingly favored competing configurations rather than a coherent retained B state. These calculations were used to define the qualitative design window before the real-MOF search, not as a material prediction in their own right.

#### *S5.5.2 Chemically modified MOF-801-derived models*

After the artificial-model stage, an explicit chemical-variant library was constructed from the ordered experimental MOF-801 parent model. Atom additions, removals and substitutions were made at L12 and L13, two positions in distinct symmetry orbits relative to the selected pocket and anchor. The frozen library contained 46 modified structures and one pristine control. The paired-acetate variants V016 and V040 failed the construction geometry screen and were excluded before the simulation library was frozen.

The table below identifies the frozen variants. OH0/OH1 and Cl0/Cl1 identify which former carboxylate oxygen position carries the anionic termination and which carries coordinated water. These were constructed rigid-backbone chemical models, with no force-field scaling. The artificial model instead retained the parent geometry while modifying host charges and scaling a selected cross interaction; it was a separate stage.

| Modification | L12 IDs | L13 IDs |
|---|---|---|
| Pristine control | V000_pristine | — |
| Missing fumarate; formate/formate terminations | V001 | V025 |
| Missing fumarate; OH/$H_2O$ terminations, assignments 00, 01, 10, 11 | V002–V005 | V026–V029 |
| Missing fumarate; Cl/$H_2O$ terminations, assignments 00, 01, 10, 11 | V006–V009 | V030–V033 |
| Formate/OH0, formate/OH1, OH0/formate, OH1/formate | V010–V013 | V034–V037 |
| Formate/acetate and acetate/formate | V014–V015 | V038–V039 |
| Single methyl substitution at alkene end 0 or 1 | V017–V018 | V041–V042 |
| Single chloro substitution at end 0 or 1 | V019–V020 | V043–V044 |
| Single bromo substitution at end 0 or 1 | V021–V022 | V045–V046 |
| Single iodo substitution at end 0 or 1 | V023–V024 | V047–V048 |

All 752 dry tasks completed successfully: 188 pose searches and 564 canonical chains. All 47 models received completed dry analyses, but none passed the combined qualification criteria, and no hydration calculations were launched for this library. All 47 passed the spatial-localization check. Competing state C exceeded 50% in 45 models. V026, V029, V032 and V036 passed the B-retention checklist but had dominant C populations and reversed dry A/B preference. V010 and V011 retained substantial A populations without establishing the required independent B minimum/projected basin.

The decision to broaden the material search therefore followed failure of the combined dry qualification criteria. These results do not establish failure under hydration or the behavior of fully relaxed or synthesized MOF-801 derivatives.

Archive records: mc010_pilot_v1/mof801_chemical_v1/manifests/variant_library.csv and library_freeze.json; scripts/build_library.py; structures/candidates/<variant>/structure.json and the associated CIFs; results/dry/*.json; results/campaign_table.csv; and raw/dry/<variant>/ trajectories and completion receipts. The separate artificial-model definition is in mc010_pilot_v1/mof801_toy_v1/manifests/final_model_definition.json.

### S5.6 CoRE MOF screening procedure

The MC-010 and MC-015 campaigns used the same locally archived 12,093-entry CoRE working corpus, with matching identifiers and recorded CIF hashes. It contained 6,714 ASR, 4,825 FSR and 554 ION entries. The upstream local source inventory contained 12,499 entries (6,963 ASR, 4,978 FSR and 558 ION); 406 previously frozen exclusions had already been applied before either campaign. The local manifests and hashes identify this working snapshot, but do not establish an exact upstream public-release tag or download date.

MC-010 stage-one filtering excluded 1,390 entries by descriptor criteria: 998 by the largest-cavity-diameter (LCD) criterion alone, 189 by both LCD and pore-limiting-diameter (PLD) criteria, 178 by PLD alone, 20 as extremely open, four for lacking the required carbon/polar-element combination, and one as extremely open with unresolved occupancy. Three additional entries were technically deferred: S01094, 2007[Ag][nan]3[ION]1, for severe source heavy-atom overlap; S08064, 2017[Zr][csq]3[FSR]1, for unresolved occupancy; and S08519, 2018[ZnMo][qtz]3[ION]2, for missing/nonneutral charges and unverified charge provenance. Thus 12,093 − 1,390 − 3 = 10,700 entries were retained.

All 10,700 retained entries completed static geometric/energetic screening. This completion count does not mean that every entry underwent dry-chain or hydration validation. Rather than exhaustively hydrating every pocket and loading, the discovery stage evaluated one primary pocket per candidate MOF with short trajectories from A- and B-like starting orientations. Candidates were promoted adaptively for validation on the basis of saved dry and hydrated diagnostics rather than after completion of the entire corpus.

The shared charged archive, cifs.tar.gz, has recorded SHA-256 e1442863da3e93ffecd75b02c8d1fb05dba91d2133ec1f344833b52037c92d8b. This hash appears in mc010_core_screen_v1/manifests/source_verification.json and mc015_quantum_site_inversion_v1/full_corpus_v4/manifest.json. Snapshot lineage is retained in assets/large_core_registry_v1/REGISTRY.json (study archive), provenance/source_database_manifest.sha256 and CIF_MANIFEST.sha256. MC-010 dispositions and filter settings are recorded in mc010_core_screen_v1/manifests/stage1_all_structures.csv and protocol/screening.json; operations_v7/summary.json records static_complete = static_target = 10700.

### S5.7 Candidate validation and replicate trajectories

The completed matched-validation campaign contained 291 selected leads. Each lead was represented by eight matched finite chains: dry and hydrated conditions, each from A and B starts, with two independent seeds per start. The principal hydrated validation condition used four local water molecules. New matched chains used 5,000 burn-in sweeps followed by 5,000 production sweeps; previously verified trajectories were reused when they satisfied the same saved condition definitions. Across the completed campaign, 2,258 validation chains were newly generated and 70 verified existing chains were reused, giving 2,328 matched chains in total. Completion of this protocol establishes finite-run reproducibility under the sampled conditions; it does not by itself establish equilibrium free energies, the many-ethanol collective extension of MC-010 or experimental material suitability.

The recorded matched-validation classifications make the campaign yield explicit. Of the 291 leads, 8 (2.7%) were classified as reproducible A-to-B-like, 151 (51.9%) as strong partial phenomena, 79 (27.1%) as bistable or initialization-dependent, and 53 (18.2%) as not reproduced or showing no useful signal. These are descriptive finite-chain categories under the campaign criteria, not equilibrium-state assignments or estimates of the prevalence of MC-010 across CoRE.

Production frames were classified using the frozen A/B/C definitions while candidate-level diagnostics retained individual start and seed identities. In addition to pooled A/B/C populations, the analysis tracked localization or outside-pocket displacement, spatial coherence of B, production-half drift, independent-seed spread and framework–ethanol and ethanol–water contacts relevant to the proposed mechanism. Contradictory seeds were retained rather than reconciled or discarded. Candidate labels in the validation reports summarize these finite-chain diagnostics and were not interpreted as direct measurements of equilibrium basin probabilities.

Structural source note. CoRE MOF 2024 identifies 2019[Zn][sqp]3[FSR]1 as the publisher-SI-derived structure c7dt04484d2_1 (DOI 10.1039/C7DT04484D). The CoRE SI metadata provides no CSD refcode for this entry. For case-study structures without a verified CSD refcode in the available structural metadata, the exact CoRE identifier is therefore used rather than assigning a refcode by inference.

The primary real-MOF example, 2019[Zn][sqp]3[FSR]1, had pooled dry populations of approximately 45.2% A, 5.2% B and 49.6% C. Under the principal four-water hydrated condition, the pooled distribution shifted to approximately 0% A, 97.2% B and 2.8% C. The four hydrated validation chains, spanning A- and B-like starts and two independent seeds per start, each strongly populated B (approximately 93–99%) while retaining ethanol in the intended local pocket. A second material, 2012[Cu][rtl]3[FSR]2, showed the same qualitative direction with smaller enrichment: its dry ensemble contained approximately 46.2% A and 1.9% B, whereas hydration increased B to approximately 48.3%. These are finite-run production-frame fractions under the saved validation conditions rather than equilibrium probabilities.

### S5.8 Additional partial realizations and negative results

After completion of the 291-lead campaign, new candidates were compared with the historical MC-010 finalists at the same evidentiary depth. This comparison reused 88 existing primary chains and added 40 missing short loading checks needed to match the earlier coverage; no free-energy calculation, enhanced sampling or stronger methodological stage was introduced. Additional fixed-loading checks on selected finalists were therefore treated as secondary tests rather than as part of the primary matched-validation classification.

For 2019[Zn][sqp]3[FSR]1, single-start fixed-loading follow-ups at six and eight local water molecules did not reproduce the strongly B-rich state observed under the principal four-water condition. These secondary checks therefore support a finite hydration window within the tested model rather than a monotonic increase of B occupancy with increasing water loading.

### S5.9 MC-010 simulation parameters and analysis definitions

MC-010 used TraPPE-UA ethanol (Ref. S1) and SPC/E water (Ref. S2). Framework Lennard-Jones interactions used DREIDING parameters for nonmetal framework atoms (Ref. S11) and UFF parameters for metal atoms (Ref. S4). Framework electrostatics used PACMAN-predicted DDEC6 partial charges supplied with the working CoRE MOF 2024 structures (Ref. S5).

Electrostatics were stage-specific. In the later custom host engines, the periodic framework potential acted on both ethanol and mobile water. The real-space Ewald term used a 12 Å cutoff and a splitting parameter of 0.2650583036 $Å^{-1}$; the reciprocal host potential was interpolated during sampling. Guest–guest Coulomb interactions within the finite local cluster used uncut direct Cartesian separations, without minimum-image wrapping or guest-image sums. The direct guest–guest treatment applied to the host-containing hydrated calculations as well as to the framework-free clusters. The early Zn-MOF-74 RASPA and custom potential-of-mean-force (PMF) calculations retained different guest-image contributions, as specified below.

| Executed stage | Boundary conditions | Framework–guest electrostatics | Guest–guest and guest-image treatment |
|---|---|---|---|
| Framework-free ethanol–water clusters | Finite cluster with positional constraints | No framework | Direct full-range Coulomb; no Ewald, cutoff or periodic images |
| Early Zn-MOF-74 RASPA dry pilot | Periodic 4 × 4 × 4 supercell | Ewald; saved precision $10^{-6}$ and 12 Å cutoff | Periodic adsorbate electrostatics, including image contributions |
| Early Zn-MOF-74 custom dry PMF | Local restrained pocket in a periodic host | Real-space plus explicit reciprocal host sum | Explicit periodic ethanol self-image/reaction correction represented by a saved polynomial |
| Later canonical-host dry calculations, including MOF-801 | One ethanol in a rigid periodic host | Periodic host Ewald potential | No intermolecular guest pairs; ethanol self-images omitted |
| Artificial symmetry-broken MOF-801, dry/hydrated | Periodic modified host; finite local water region when hydrated | Host Ewald with deliberately modified host charges | Direct finite-cluster guest pairs in hydrated runs; no guest images |
| Chemical MOF-801 library | Periodic rigid chemical models; dry runs only | Periodic host Ewald potential | No mobile guest–guest pairs; fixed cap waters are part of the host |

| Executed stage | Boundary conditions | Framework–guest electrostatics | Guest–guest and guest-image treatment |
|---|---|---|---|
| Real-CoRE dry/hydrated campaign | Periodic rigid host; finite local water cluster when hydrated | Host Ewald for ethanol and water | Dry: no intermolecular guest pairs. Hydrated: direct ethanol–water and water–water Coulomb; no pair cutoff, wrapping or guest-image sums |

The stage assignments follow saved run inputs, completion records and the active energy-evaluation paths. Framework-free calculations used mc010_pilot_v1/no_mof_orientation_v1/engine/orientation_mc_v2.cpp and molecules.hpp. The early RASPA input is mc010_pilot_v1/raw/dry_01/ZnMOF74_T300_s101031/simulation.input; the early custom PMF used mc010_pilot_v1/dry_orientation_v1/engine/dry_pmf.cpp, with a nonzero periodic_self_K term retained in raw/pmf_seed1/samples.csv.

Later canonical-host calculations used mc010_pilot_v1/intrinsic_pockets_v1/engine/dry_mc.cpp and host.cpp; the artificial model used mc010_pilot_v1/mof801_toy_v1/engine/experiment.cpp. CoRE production was launched by mc010_core_screen_v1/scripts/run_chain.py and used engine/mc.cpp with guest pairs evaluated in engine/molecules.hpp. Archived inputs and receipts for the primary lead are retained in mc010_core_screen_v1/figures/S09559_structural_v1/sources/trajectories. The complete parameter files, Lennard-Jones definitions, mixing conventions and campaign-specific settings remain in the study archive.

The matched real-MOF validation used the frozen state definitions described in Section S5.2.

New validation chains comprised 5,000 burn-in sweeps followed by 5,000 production sweeps.

The principal hydrated condition contained four local water molecules, and each selected lead was evaluated under dry and hydrated conditions from A- and B-start configurations with two independent seeds per start. Production-frame fractions, localization, seed spread, production-half drift and mechanism-relevant framework–ethanol and ethanol–water contacts were retained as diagnostics rather than interpreted as exact equilibrium probabilities.

Structural illustrations were generated from saved classified trajectory frames without coordinate optimization. Representative dry-A and hydrated-B configurations were selected from their respective ensembles using the archived frame-selection procedure rather than by visual preference. Paired dry and hydrated images therefore illustrate typical configurations from separate chains and are not presented as recorded transition sequences.

## S6. Supplementary information for MC-015

### S6.1 Original generated hypothesis and provenance

MC-015 maps to the single prospective proposal PV0-031, generated from G-U02 molecular hydrogen, H06 connected spatially heterogeneous pore environments and the cryogenic adsorptive-storage context. The guest card described $H_2$ as a very small, light rigid diatomic for which translational and rotational quantum effects can matter at low temperature. The host card specified two or more connected pore regions that differ in size, shape or wall chemistry and explicitly did not assign an occupancy order or migration path. The operating context specified pure $H_2$ under cryogenic loading but no numerical temperature window, dilute-regime inversion target or named material.

The common instruction asked for one unusual qualitative phenomenon, its physical reasoning, distinguishing signature, relevant conditions, application relevance, falsifier and required host characteristics, or an abstention. The archived request contained no tools, retrieval, previous response or specific-MOF search. The response introduced the deep/sharp–shallow/soft competition and named the behavior 'Quantum-induced inverse thermal partitioning': a classically deeper region can incur a larger confinement penalty than a shallower region, causing the population preference to reverse as quantum effects strengthen. It further proposed the matched classical control, low-loading comparison and equilibration falsifier. Supplementary Table S5 summarizes the boundary between supplied information, generated relationships and later implementation choices.

The recorded call used gpt-6-astra with reasoning effort max and text verbosity medium on 28 September 2026. One attempt completed without retries. The request specified tools = [], tool_choice = none and no previous_response_id; temperature, top_p and seed were not explicitly supplied.

### S6.2 Hydrogen interaction model and Feynman–Hibbs treatment

MC-015 used rigid framework structures, neutral single-site Buch $H_2$ (Ref. S3), UFF framework van der Waals parameters (Ref. S4) and a quadratic Feynman–Hibbs (FH) correction. Because the Buch $H_2$ site is neutral, framework partial charges do not contribute to the $H_2$–framework interaction energy in this model. Classical and FH calculations retained the same structural inputs and underlying interaction parameters; the comparison changed the semiclassical treatment rather than the chemical model.

The FH calculation was used as a bounded semiclassical test of the generated regional-preference signature rather than as a complete translational-rotational quantum treatment. The warmer endpoint is not assumed to be a classical limit because the correction-off calculation at the same temperature supplies the matched classical reference. Relevant prior discussions of nuclear quantum effects for $H_2$ adsorption in MOFs are given in Refs. S9 and S10.

### S6.3 Numerical implementation and benchmark calculations

The analytic regional-integration implementation was benchmarked against direct RASPA Widom calculations (Ref. S8) for IRMOF-1 using the same $H_2$ interaction model. At 77 K, the classical and FH Henry coefficients were 7.37765 × 10^-5 and 5.82968 × 10^-5 mol kg^-1 Pa^-1, respectively. The grid and Widom values differed by 0.0373% for the classical model and 0.0170% for FH, and the FH/classical ratios from the two routes differed by 0.0203%. Grid refinement changed either integrated result by less than 0.002%.

The installed native FH-grid route returned zero values in a preliminary implementation check and was therefore not used for production calculations; production used the independently checked analytic evaluator. A separate dilute-slope reproduction was retained as a consistency check rather than an uncertainty estimate or validation of the 25 K regime.

For the reported regional ratios, fixed basin masks defined from the classical site analysis were used for both the classical and FH evaluations. Within each mask, the configurational integral was evaluated by voxel quadrature, summing $\exp[-U(r)/(RT)]\Delta V$. When a region represented a symmetry-related family, its integrated weight was divided by the family multiplicity before forming the A/B ratio. Thus the reported quantity is a per-site-family regional statistical-weight ratio, not a ratio of the two local energies shown in Fig. 4a,b of the main text. For the primary R2/R7 pair of 2019[Zn][nan]3[FSR]24, both multiplicities are four and therefore cancel exactly.

### S6.4 Regional assignment and Henry-weight integration

Low-energy minima and surrounding adsorption regions were identified in the classical site analysis while preserving periodic spatial identities. The resulting basin masks were then held fixed across the classical and FH calculations and across temperature; A and B therefore denote spatial regions rather than whichever region ranks first at a given condition. Repeated or symmetry-related sites were tracked as families.

For model M and region $\Omega_i$, the configurational weight was evaluated as $Z_i^M(T) = \Sigma_{r \in \Omega_i} \exp[-U_M(r)/(RT)]\, \Delta V$. A common normalization converts these weights into regional Henry contributions, so for two regions at the same temperature their normalized regional-weight ratio equals the corresponding dilute equilibrium population ratio. For a ratio $r = Z_A/Z_B$, the A fraction within A+B is $r/(1+r)$; the contribution of A+B to total adsorption was retained as a separate quantity.

For site families containing $m_A$ and $m_B$ symmetry-equivalent regions, the comparison used $(Z_A/m_A)/(Z_B/m_B)$. Family totals, multiplicities and within-pair fractions were therefore kept distinct. Summing all assigned contributions plus the unassigned remainder checks conservation of the total integral but does not by itself validate the physical placement of basin boundaries.

### S6.5 Full-corpus screening procedure and exclusion accounting

The final compatibility pass used the shared 12,093-entry working snapshot described in Section S5.6. The project-specific coordinate fingerprint grouped entries into 8,887 records by hashing the cell parameters and sorted element/fractional-coordinate/occupancy tuples; framework charges were excluded from this fingerprint. This grouping defines coordinate uniqueness for the present campaign. Parameter coverage, nonorthogonal-cell support, geometry, size, quality and metadata guards were applied without repairing excluded structures. Energy fields were evaluated for 456 eligible structures, and 174 yielded usable regional assignments. Supplementary Table S6 gives the full screening funnel.

Before grid generation, 4,924 coordinate-unique structures lacked parameter coverage, 3,244 used unsupported nonorthogonal cells, 126 failed geometry guards, 96 exceeded size guards, 33 failed quality checks and eight lacked required metadata. A further 282 grid-calculated structures did not yield usable regional assignments. These are implementation exclusions, not evidence that the omitted structures cannot display MC-015.

Random exploratory batches preceded the final corpus-wide pass, and previously generated fields were reused. The final completion pass added 54 grids and 20 usable assignments in 76 s; this incremental time is not the total computational cost of method development, validation and screening.

### S6.6 Strong-candidate definition and screening results

The discovery search ranked temperature-dependent crossings between non-equivalent regions with appreciable statistical weight and compared their classical and FH responses. It identified 53 candidate crossover pairs in 26 structures; seven pairs in four structures received the pipeline's stronger-candidate label. These counts derive from approximate regional assignments and are not 53 independently validated phenomena. The operational threshold for the stronger label was fixed in the selection code before finalist characterization and was not redefined from the final results.

Only three leads received the additional checked characterization used in the main text. Unsupported structures and ambiguous assignments were not scored as negatives, so the screening yield should not be interpreted as an estimate of MC-015 prevalence across CoRE.

### S6.7 Detailed characterization of the selected leads

The three checked records do not represent three independent frameworks. The 2019 FSR record m157 and the 2019 ASR record n140 use the same underlying host coordinates. Their formulas are $Zn_8H_{60}C_{56}N_4O_{32}$ and $Zn_8H_{32}C_{44}O_{28}$, respectively; the difference, $C_{12}H_{28}N_4O_4$, corresponds exactly to four DMF molecules. Thus the FSR input retains four DMF molecules that are absent from the ASR input. The 2014 record n025 provides the distinct supporting framework.

The primary checked lead was 2019[Zn][nan]3[FSR]24 (analysis ID m157), using regions R2 and R7 as A and B; the two site families each have multiplicity four. At the representative A and B positions marked in Fig. 4a,b of the main text, the 25 K local difference $\Delta U_{B-A} = U_B - U_A$ changed from +0.50 kJ mol$^{-1}$ in the classical field to −0.22 kJ mol$^{-1}$ with the FH correction. These are point-energy values and are distinct from the integrated regional weights. On the saved finer-grid fields, the FH A/B ratio decreased from 1.520 at 100 K to 0.595 at 25 K and crossed unity near 35 K, whereas the classical ratio increased from 1.660 to 4.541. Supplementary Table S7 reports the endpoint ratios and within-pair population fractions.

The related 2019[Zn][nan]3[ASR]25 record (analysis ID n140) used regions R2 and R3; both site families have multiplicity four. Its finer-grid FH ratio changed from approximately 1.70 at 100 K to 0.51 at 25 K and crossed near 42 K. Within the pair this corresponds to approximately 63% A at 100 K and 34% A at 25 K. The classical response was weak and nonmonotonic, so this case supports the FH crossover but not the clearer opposite classical trend of 2019[Zn][nan]3[FSR]24.

The 2014[Zn][nan]3[ASR]10 record (analysis ID n025) used regions R1 and R3, both with multiplicity eight, and was compared from 45 to 25 K. The finer-grid FH crossing occurred near 32 K. Region A retained the deeper FH point minimum even when B acquired the larger integrated population, emphasizing that basin volume contributes to the regional free energy. This case showed greater numerical sensitivity and was retained as supporting evidence. Supplementary Table S8 summarizes the three checked leads.

### S6.8 Structural interpretation of the competing adsorption regions

For each checked pair, the FH penalty evaluated at the classical minimum of A was larger than the corresponding penalty at B, consistent with the generated deep/sharp versus shallow/soft mechanism. However, three previously processed non-switchers also exhibited a larger local penalty in A. A differential local penalty is therefore not sufficient: the initial classical preference and the distribution of statistical weight throughout each region also control whether a crossover occurs.

The 2014 supporting example illustrates this point directly because its population inversion occurs without an exchange in the ranking of the corrected point minima. The integrated regional observable therefore contains information that an isolated minimum-energy comparison would miss.

### S6.9 Numerical sensitivity and model limitations

The regional assignments were generated by a deliberately inexpensive analysis. A general-purpose IRMOF-1 prototype showed that family populations could be sensitive to a half-voxel shift of the grid origin even when total Henry coefficients were stable. Screening assignments were therefore treated as approximate and additional checks were reserved for selected leads.

The two 2019 checked records use the same underlying host coordinates but differ by four retained DMF molecules in the FSR input ($C_{12}H_{28}N_4O_4$ in total, 48 atoms). They are therefore related structural inputs rather than independent framework families. Because the primary m157 result is obtained for the DMF-containing FSR structure, its result is interpreted for that specified structural input rather than attributed to the solvent-free host alone.

Finally, the quadratic FH treatment is a semiclassical approximation at low temperature, and the reported crossover temperatures and amplitudes should not be interpreted as precise experimental predictions. The regional integral describes an equilibrium population ratio for the fixed model but not an exchange timescale. No new NVT, GCMC or explicit quantum trajectories were added for the finalist characterization.

## S7. Generated mechanisms not advanced as principal case studies

### S7.1 MC-004

MC-004 proposed that increasing hydration could eventually reduce the absolute number of water-framework hydrogen bonds even while total water loading continued to increase because the adsorbed water reorganizes into a more favorable water-water network. The calculation itself was straightforward and received the highest executability tier. Closer inspection of the literature, however, showed that changes in water-MOF and water-water hydrogen bonding with loading, confined-water network growth and their connection to adsorption and dynamics are already widely studied. The remaining distinction, an actual decrease in the absolute number of surface contacts rather than only a declining fraction, was too narrow to justify a major case study here. MC-004 was therefore deprioritized on novelty grounds rather than because it was computationally inaccessible.

### S7.2 MC-020

MC-020 proposed that methanol molecules in a cavity next to a narrow rigid aperture can form hydrogen-bonded arrangements incompatible with the orientation needed for escape. Successful passage would then require neighboring guests to reorient and rebond, creating a transient gate controlled by the adsorbates rather than by framework motion.

The material-search stage retained 3,856 cavity-neck candidates and a preferred chemistry/model-ready pool of 1,342 structures. The executed screening used the TraPPE-UA methanol model (Ref. S1). The decisive criterion was not low diffusivity or a loading-dependent barrier alone, but event-level coupling between successful passage and neighboring reorientation/rebonding together with an isolated-guest control.

The independent ZIF-8 methanol benchmark remained unresolved because of sampling diagnostics. Accordingly, MC-020 was not treated as a positive realization. It was deferred on proof burden, and only its stopping logic and implemented molecular-model provenance are retained here.

### S7.3 MC-025

MC-025 proposed cooperative exchange of bound ethanol by water. In this mechanism, replacement of one ethanol molecule creates local water connectivity that makes replacement of neighboring bound ethanol more favorable. The defining feature is therefore neighbor-dependent site exchange rather than ordinary competitive water adsorption.

The first CoRE material search exposed a practical limitation. Although repeated oxygen- or metal-containing environments were common, structural descriptors alone could not establish a shared reversible site that plausibly binds both ethanol and water or the intersite geometry needed for cooperative water bridging. Twenty structural leads were reviewed without obtaining a defensible mechanism-focused shortlist. Substantially more chemistry and site-access analysis would be required before the decisive mixture calculation could be attempted.

MC-025 would also have added another ethanol/water hydrogen-bonding case to a manuscript already containing MC-010. Because the host could not be identified cleanly and a positive result would add limited physical breadth, it was not pursued further in this study.

### S7.4 Other retained mechanism families

The status of MC-006, MC-018, MC-019 and MC-022 is described in Section S4.7 and summarized in Table S4.

## S8. Supplementary tables

### Table S1. Application and operating contexts

Table S1. Complete application and operating-context inventory. The application and operating text are reproduced from the frozen records. PV0-001–020 retain their original descriptions. The iodine entry was added from the existing guest card when the matrix was completed. The context keys link this inventory to Table S2. The source cases are provenance labels and were not an additional model input.

| Context key / source / guest | Application | Operating context supplied |
|---|---|---|
| C001<br>PV0-001<br>G-U01 | $CO_2$ capture and sorbent regeneration | Pure dry $CO_2$; near-ambient gas dosing followed by pressure reduction; temperature, loading, and regeneration history recorded. |
| C002<br>PV0-002<br>G-U01 | $CO_2$ storage and pressure buffering | Pure dry $CO_2$; controlled pressure cycles at fixed near-ambient temperature; activated starting state and equilibration time recorded. |
| C003<br>PV0-003<br>G-U02 | Cryogenic adsorptive hydrogen storage | Pure molecular $H_2$; cryogenic gas loading over a pressure range; temperature, accessible volume, and uptake/release times recorded. |
| C004<br>PV0-004<br>G-U02 | Heat management in a cryogenic hydrogen sorbent | Pure molecular $H_2$; cryogenic loading at controlled temperature and pressure; thermal measurements at defined loading and sample geometry. |
| C005<br>PV0-005<br>G-U03 | Water-based adsorption heating and cooling | Pure water vapor; sub-saturation vapor-pressure cycles with adsorption and regeneration temperatures recorded; hydrolytically compatible host required. |
| C006<br>PV0-006<br>G-U03 | Humidity capture and water storage | Pure water vapor as a simplified humidity input; relative pressure swept below bulk saturation near ambient temperature; loading history and connected pore volumes recorded. |
| C007<br>PV0-007<br>G-U04 | Methanol vapor recovery and sorbent regeneration | Pure methanol vapor; controlled sub-saturation vapor pressure near ambient temperature followed by regeneration; temperature and observation time recorded. |
| C008<br>PV0-008<br>G-U04 | Heat management in a methanol adsorption heat exchanger | Pure methanol vapor; controlled loading below bulk saturation at a stated operating temperature; thermal measurement geometry and empty-host reference recorded. |

| Context key / source / guest | Application | Operating context supplied |
|---|---|---|
| C009<br>PV0-009<br>G-B01 | $CO_2$ removal from an idealized dry flue-gas pair | Dry $N_2$-rich $CO_2$/$N_2$ gas near ambient temperature and low total pressure; feed fraction, pressure, residence time, and uptake history specified later. |
| C010<br>PV0-010<br>G-B01 | $CO_2$ removal from nitrogen-rich process gas | Dry $N_2$-rich $CO_2$/$N_2$ gas near ambient temperature; low $CO_2$ partial pressure with feed composition and regeneration conditions recorded. |
| C011<br>PV0-011<br>G-B01 | $CO_2$ capture from an idealized dry flue-gas pair | Dry $N_2$-rich $CO_2$/$N_2$ gas near ambient temperature and low total pressure; composition, residence time, and equilibration protocol controlled. |
| C012<br>PV0-012<br>G-B02 | Idealized biogas upgrading | Dry binary $CO_2$/$CH_4$ gas near ambient temperature; feed ratio and total pressure controlled; no humidity or higher hydrocarbons included. |
| C013<br>PV0-013<br>G-B02 | $CO_2$ removal from methane-rich natural gas | Dry $CH_4$-rich $CO_2$/$CH_4$ gas under elevated pressure near ambient temperature; compression/decompression history and starting host phase recorded. |
| C014<br>PV0-014<br>G-B03 | Ethylene purification | Dry binary $C_2H_4$/$C_2H_6$ gas above its bulk dew point; controlled temperature, total pressure, composition, and contact time. |
| C015<br>PV0-015<br>G-B03 | Ethylene purification | Dry binary $C_2H_4$/$C_2H_6$ gas above its bulk dew point; controlled temperature, pressure, composition, and site activation state. |
| C016<br>PV0-016<br>G-B03 | $C_2$ hydrocarbon recycle separation | Dry binary $C_2H_4$/$C_2H_6$ gas above its bulk dew point; temperature, feed ratio, contact time, and prior activation recorded. |
| C017<br>PV0-017<br>G-B04 | Vapor-phase ethanol dehydration | Ethanol-rich binary ethanol/water vapor; heated operation above the mixture dew point; partial pressures and regeneration history controlled. |
| C018<br>PV0-018<br>G-B04 | Ethanol recovery from water-containing solvent vapor | Water-rich binary ethanol/water vapor above the mixture dew point; feed partial pressures, temperature, and contact time controlled. |
| C019<br>PV0-019<br>G-B05 | para-Xylene enrichment | Binary para-/ortho-xylene vapor in a heated single-vapor-phase feed; composition, temperature, partial pressures, and residence time controlled. |
| C020<br>PV0-020<br>G-B05 | Binary aromatic-isomer fractionation | Binary para-/ortho-xylene vapor in a heated single-vapor-phase feed; composition, contact time, thermal history, and regeneration conditions recorded. |
| CI2<br>Extension only<br>G-U05 | iodine-vapor capture | Molecular iodine ($I_2$) vapor as the sole incoming guest species; iodine-vapor capture. No temperature, pressure, or exposure-time regime is specified in the frozen guest card. |

## Table S2. Complete 80-case input-output map

Table S2. Complete 80-case guest–host–context and outcome map. Context keys correspond to Table S1. Family labels were assigned after generation and were not model inputs.

| Case | Batch | Guest | Host | Context | Outcome | Family |
|---|---|---|---|---|---|---|
| PV0-001 | Initial 20 | G-U01 | H02 | C001 | ABSTAIN | Not assigned |
| PV0-002 | Initial 20 | G-U01 | H05 | C002 | Proposal | MC-001 |
| PV0-003 | Initial 20 | G-U02 | H01 | C003 | Proposal | MC-002 |
| PV0-004 | Initial 20 | G-U02 | H07 | C004 | Proposal | MC-003 |
| PV0-005 | Initial 20 | G-U03 | H03 | C005 | Proposal | MC-004 |
| PV0-006 | Initial 20 | G-U03 | H06 | C006 | Proposal | MC-005 |
| PV0-007 | Initial 20 | G-U04 | H04 | C007 | Proposal | MC-006 |
| PV0-008 | Initial 20 | G-U04 | H07 | C008 | Proposal | MC-003 |
| PV0-009 | Initial 20 | G-B01 | H01 | C009 | ABSTAIN | Not assigned |
| PV0-010 | Initial 20 | G-B01 | H02 | C010 | ABSTAIN | Not assigned |
| PV0-011 | Initial 20 | G-B01 | H06 | C011 | Proposal | MC-007 |
| PV0-012 | Initial 20 | G-B02 | H03 | C012 | ABSTAIN | Not assigned |
| PV0-013 | Initial 20 | G-B02 | H05 | C013 | Proposal | MC-008 |
| PV0-014 | Initial 20 | G-B03 | H01 | C014 | ABSTAIN | Not assigned |
| PV0-015 | Initial 20 | G-B03 | H02 | C015 | ABSTAIN | Not assigned |
| PV0-016 | Initial 20 | G-B03 | H04 | C016 | Proposal | MC-009 |
| PV0-017 | Initial 20 | G-B04 | H03 | C017 | Proposal | MC-010 |
| PV0-018 | Initial 20 | G-B04 | H06 | C018 | Proposal | MC-011 |
| PV0-019 | Initial 20 | G-B05 | H01 | C019 | ABSTAIN | Not assigned |
| PV0-020 | Initial 20 | G-B05 | H04 | C020 | Proposal | MC-009 |
| PV0-021 | Extension 60 | G-U01 | H01 | C001 | Proposal | MC-012 |
| PV0-022 | Extension 60 | G-U01 | H03 | C001 | ABSTAIN | Not assigned |
| PV0-023 | Extension 60 | G-U01 | H04 | C001 | Proposal | MC-006 |
| PV0-024 | Extension 60 | G-U01 | H06 | C001 | ABSTAIN | Not assigned |

| Case | Batch | Guest | Host | Context | Outcome | Family |
|---|---|---|---|---|---|---|
| PV0-025 | Extension 60 | G-U01 | H07 | C001 | Proposal | MC-003 |
| PV0-026 | Extension 60 | G-U01 | H08 | C001 | ABSTAIN | Not assigned |
| PV0-027 | Extension 60 | G-U02 | H02 | C003 | Proposal | MC-013 |
| PV0-028 | Extension 60 | G-U02 | H03 | C003 | ABSTAIN | Not assigned |
| PV0-029 | Extension 60 | G-U02 | H04 | C003 | Proposal | MC-014 |
| PV0-030 | Extension 60 | G-U02 | H05 | C003 | Proposal | MC-014 |
| PV0-031 | Extension 60 | G-U02 | H06 | C003 | Proposal | MC-015 |
| PV0-032 | Extension 60 | G-U02 | H08 | C003 | Proposal | MC-016 |
| PV0-033 | Extension 60 | G-U03 | H01 | C005 | Proposal | MC-017 |
| PV0-034 | Extension 60 | G-U03 | H02 | C005 | ABSTAIN | Not assigned |
| PV0-035 | Extension 60 | G-U03 | H04 | C005 | Proposal | MC-018 |
| PV0-036 | Extension 60 | G-U03 | H05 | C005 | Proposal | MC-001 |
| PV0-037 | Extension 60 | G-U03 | H07 | C005 | Proposal | MC-019 |
| PV0-038 | Extension 60 | G-U03 | H08 | C005 | Proposal | MC-017 |
| PV0-039 | Extension 60 | G-U04 | H01 | C007 | Proposal | MC-020 |
| PV0-040 | Extension 60 | G-U04 | H02 | C007 | ABSTAIN | Not assigned |
| PV0-041 | Extension 60 | G-U04 | H03 | C007 | Proposal | MC-004 |
| PV0-042 | Extension 60 | G-U04 | H05 | C007 | Proposal | MC-001 |
| PV0-043 | Extension 60 | G-U04 | H06 | C007 | ABSTAIN | Not assigned |
| PV0-044 | Extension 60 | G-U04 | H08 | C007 | Proposal | MC-011 |
| PV0-045 | Extension 60 | G-U05 | H01 | CI2 | ABSTAIN | Not assigned |
| PV0-046 | Extension 60 | G-U05 | H02 | CI2 | Proposal | MC-021 |
| PV0-047 | Extension 60 | G-U05 | H03 | CI2 | ABSTAIN | Not assigned |
| PV0-048 | Extension 60 | G-U05 | H04 | CI2 | Proposal | MC-006 |
| PV0-049 | Extension 60 | G-U05 | H05 | CI2 | Proposal | MC-001 |
| PV0-050 | Extension 60 | G-U05 | H06 | CI2 | Proposal | MC-011 |
| PV0-051 | Extension 60 | G-U05 | H07 | CI2 | Proposal | MC-003 |

| Case | Batch | Guest | Host | Context | Outcome | Family |
|---|---|---|---|---|---|---|
| PV0-052 | Extension 60 | G-U05 | H08 | CI2 | Proposal | MC-011 |
| PV0-053 | Extension 60 | G-B01 | H03 | C009 | ABSTAIN | Not assigned |
| PV0-054 | Extension 60 | G-B01 | H04 | C009 | Proposal | MC-022 |
| PV0-055 | Extension 60 | G-B01 | H05 | C009 | Proposal | MC-023 |
| PV0-056 | Extension 60 | G-B01 | H07 | C009 | Proposal | MC-003 |
| PV0-057 | Extension 60 | G-B01 | H08 | C009 | Proposal | MC-007 |
| PV0-058 | Extension 60 | G-B02 | H01 | C012 | Proposal | MC-007 |
| PV0-059 | Extension 60 | G-B02 | H02 | C012 | ABSTAIN | Not assigned |
| PV0-060 | Extension 60 | G-B02 | H04 | C012 | Proposal | MC-006 |
| PV0-061 | Extension 60 | G-B02 | H06 | C012 | Proposal | MC-007 |
| PV0-062 | Extension 60 | G-B02 | H07 | C012 | Proposal | MC-003 |
| PV0-063 | Extension 60 | G-B02 | H08 | C012 | Proposal | MC-007 |
| PV0-064 | Extension 60 | G-B03 | H03 | C014 | ABSTAIN | Not assigned |
| PV0-065 | Extension 60 | G-B03 | H05 | C014 | Proposal | MC-024 |
| PV0-066 | Extension 60 | G-B03 | H06 | C014 | Proposal | MC-007 |
| PV0-067 | Extension 60 | G-B03 | H07 | C014 | Proposal | MC-003 |
| PV0-068 | Extension 60 | G-B03 | H08 | C014 | Proposal | MC-007 |
| PV0-069 | Extension 60 | G-B04 | H01 | C017 | Proposal | MC-012 |
| PV0-070 | Extension 60 | G-B04 | H02 | C017 | Proposal | MC-025 |
| PV0-071 | Extension 60 | G-B04 | H04 | C017 | Proposal | MC-006 |
| PV0-072 | Extension 60 | G-B04 | H05 | C017 | Proposal | MC-024 |
| PV0-073 | Extension 60 | G-B04 | H07 | C017 | Proposal | MC-026 |
| PV0-074 | Extension 60 | G-B04 | H08 | C017 | Proposal | MC-027 |
| PV0-075 | Extension 60 | G-B05 | H02 | C019 | ABSTAIN | Not assigned |
| PV0-076 | Extension 60 | G-B05 | H03 | C019 | ABSTAIN | Not assigned |
| PV0-077 | Extension 60 | G-B05 | H05 | C019 | Proposal | MC-024 |
| PV0-078 | Extension 60 | G-B05 | H06 | C019 | ABSTAIN | Not assigned |

| Case | Batch | Guest | Host | Context | Outcome | Family |
|---|---|---|---|---|---|---|
| PV0-079 | Extension 60 | G-B05 | H07 | C019 | Proposal | MC-028 |
| PV0-080 | Extension 60 | G-B05 | H08 | C019 | ABSTAIN | Not assigned |

The complete request and raw-response files are preserved separately in the study archive. The family column is included here only for traceability across later analyses.

## Table S3. Selected prospective outputs

Table S3. Selected original outputs illustrating how different host capabilities can lead to different qualitative questions from the supplied guest physics. The complete 80-case output record is given in Table S2 and the archived raw-response files.

| Guest system | Host capability | Operating context | Prospective output |
|---|---|---|---|
| $CO_2$ (PV0-001) | Accessible, chemically specific binding sites | $CO_2$ capture and regeneration | ABSTAIN: orientation-dependent binding judged conventional; no distinctive collective response justified. |
| $H_2$ (PV0-003) | Rigid confinement with restrictive apertures | Cryogenic adsorptive storage | Low-temperature crossover to tunneling-supported intercavity exchange. |
| $H_2$ (PV0-031) | Connected, spatially heterogeneous pore environments | Cryogenic adsorptive hydrogen storage | Quantum-induced inverse thermal partitioning: cooling shifts $H_2$ toward a classically less attractive region because the more attractive region imposes a larger confinement penalty (MC-015). |
| $H_2$ (PV0-004) | Compliant vibrational framework with measurable heat transport | Cryogenic heat management | Hybridization of confined-$H_2$ translational motion with a heat-carrying framework mode. |
| Ethanol / water (PV0-017) | Spatially patterned hydrogen-bonding surface | Vapor-phase ethanol dehydration | Water-triggered cooperative reorientation of retained ethanol (MC-010). |

## Table S4. Executability assessment and host specifications

Table S4. Disposition of the nine execution-A/B mechanism families after host specification. The four candidates not advanced further are retained as prospective mechanisms rather than classified as failures.

| Mechanism | Generated phenomenon / short label | Execution tier | Status in this study |
|---|---|---|---|
| MC-004 | Surface anchoring vs guest hydrogen-bond networking | A | Developed; deprioritized when novelty distinction narrowed |
| MC-006 | Adsorbate-stabilized restrictive local host gates | B | Host specification retained; not pursued further |
| MC-010 | Water-assisted polarity inversion of retained ethanol | A | Principal case study; positive partial mechanistic realization |
| MC-015 | Quantum confinement reverses inter-region $H_2$ partitioning | B | Principal case study; positive model-level realization |
| MC-018 | Water bridges collectively order mobile host groups | B | Host specification retained; not pursued further |
| MC-019 | Hydrogen-bonded water dynamically braces the host | B | Host specification retained; not pursued further |
| MC-020 | Guest reorientation and rebonding gate molecular escape | B | Developed; deferred because decisive proof burden became high |

| Mechanism | Generated phenomenon / short label | Execution tier | Status in this study |
|---|---|---|---|
| MC-022 | One guest stabilizes local gates open for another | B | Host specification retained; not pursued further |
| MC-025 | Water networking cooperatively exchanges bound site occupants | B | Developed; no defensible host realization identified for this study |

## Table S5. MC-015 hypothesis provenance and implementation boundary

Table S5. Original MC-015 input, model-generated additions and later implementation. The deep/sharp–shallow/soft competition and inverse-partitioning signature were introduced in the generated response rather than supplied in the H06 host card.

| Stage | Documented content | Provenance boundary |
|---|---|---|
| Supplied guest card | G-U02 molecular $H_2$ with low mass, weak interactions and low-temperature translational and rotational quantum effects. | No spatial population ordering, competing A/B potentials or crossover temperature. |
| Supplied host card | H06 connected regions differing in size, shape or wall chemistry, with geometry, chemistry, volumes and accessibility to establish later. | Explicitly no occupancy order or migration path and no deep/sharp–shallow/soft construction. |
| Supplied context and instruction | Cryogenic adsorptive storage with pure $H_2$ gas loading and temperature, volume and uptake or release times recorded. The instruction requests one qualitative phenomenon or abstention. | No dilute fixed-host inversion target, named MOF, numerical regime, FH model or Henry calculation. |
| Generated physical relationship | A classically deeper minimum can confine $H_2$ more sharply than a shallower competing region. The larger confinement penalty can outweigh A’s binding advantage. | Introduced in the original PV0-031 response before material identification. |

| Stage | Documented content | Provenance boundary |
|---|---|---|
| Generated signature and controls | Normalized A/B occupation decreases on cooling while a classical calculation using the same potentials and accessible volumes increasingly favors A, at low loading and without host transformation. | The opposite classical comparison, normalization and dilute-regime choice were already proposed by the model. |
| Generated conditions and falsifier | Accessible connected regions with equilibrated exchange. Inverse redistribution should survive equilibration. A return to deeper-region preference after longer equilibration argues for trapping instead. | The response is conditional and explicitly declines a numerical temperature or pressure window. |
| Later operational choices | Neutral single-site Buch $H_2$/UFF, quadratic FH, analytic grids, fixed approximate regions, regional Henry integration, CoRE hosts and adaptive temperature exploration. | These choices implement the generated regional-population question and were not specified in the original input or response. |

674 ## Table S6. MC-015 full-corpus screening funnel

675 Table S6. Completed MC-015 screening funnel. Coordinate uniqueness and eligibility are implementation categories
676 and do not imply independent framework families or physical impossibility for excluded structures.

| Stage | Count | Interpretation |
|---|---|---|
| Database entries checked | 12,093 | Complete corpus compatibility pass |
| Coordinate-unique records | 8,887 | Coordinate-deduplicated analysis units |
| Eligibility passes | 749 entries / 456 structures | Supported interaction parameters, cell geometry and input guards |
| Structures with energy grids | 456 | Includes reused calculations and 54 new grids in the final pass |
| Usable regional assignments | 174 | 282 grid-calculated structures remained unassigned or ambiguous |
| Candidate crossovers | 53 pairs / 26 structures | Approximate screening results |

| Stage | Count | Interpretation |
|---|---|---|
| Stronger screening candidates | 7 pairs / 4 structures | Pipeline ranking category; not an additional validation tier |
| Leads with additional checks | 3 | Two 2019 entries; one 2014 entry |

## Table S7. Primary MC-015 temperature-response example

Table S7. Primary MC-015 example for 2019[Zn][nan]3[FSR]24 (analysis ID m157), with A = R2 and B = R7; both regional families have multiplicity four. Population fractions refer to A+B rather than the whole MOF.

| Model | T (K) | Regional A/B ratio | A within A+B | B within A+B |
|---|---|---|---|---|
| Classical | 100 | 1.660 | 62.4% | 37.6% |
| Classical | 25 | 4.541 | 82.0% | 18.0% |
| FH | 100 | 1.520 | 60.3% | 39.7% |
| FH | 25 | 0.595 | 37.3% | 62.7% |

## Table S8. Checked MC-015 leads

Table S8. Temperature windows and roles of the three checked MC-015 leads. Crossings are approximate predictions of the model rather than precisely established physical transition temperatures.

| CoRE identifier / analysis ID | A/B regions (multiplicity) | Warm → cool | FH crossing | Role |
|---|---|---|---|---|
| 2019[Zn][nan]3[FSR]24 / m157 | R2 / R7 (4/4) | 100 → 25 K | ~35 K | Primary example with a clear opposite classical trend |
| 2019[Zn][nan]3[ASR]25 / n140 | R2 / R3 (4/4) | 100 → 25 K | ~42 K | Related ASR representation of the same underlying 2019 host; the primary FSR input retains four DMF molecules; weak and nonmonotonic classical response |
| 2014[Zn][nan]3[ASR]10 / n025 | R1 / R3 (8/8) | 45 → 25 K | ~32 K | Supporting basin-volume example with greater numerical sensitivity |

## Table S9. Three-run generation and mechanistic-breadth summary

Table S9. Generation outcomes and comparative mechanism counts for the original structured run, repeat structured run and guest-only baseline. Each run contained 80 generation calls. Broad-family and subtype counts are from the common three-run taxonomy and are separate from the original discovery-stage 28-family classification.

| Run | Calls | Proposals | Abstentions | Broad families | Subtypes |
|---|---|---|---|---|---|
| Original structured | 80 | 57 | 23 | 21 | 39 |
| Repeat structured | 80 | 63 | 17 | 21 | 41 |
| Guest-only | 80 | 80 | 0 | 8 | 21 |

## Table S10. Mechanistic breadth by guest system

Table S10. Numbers of broad mechanism families and subtypes represented within each guest system under the common three-run taxonomy. Each run contained eight generation calls per guest. Counts are not additive across guests because the same family or subtype can occur for more than one guest system.

| Guest system | Original families | Original subtypes | Repeat families | Repeat subtypes | Guest-only families | Guest-only subtypes |
|---|---|---|---|---|---|---|
| Carbon dioxide | 4 | 4 | 5 | 5 | 1 | 1 |
| Molecular hydrogen | 5 | 7 | 5 | 7 | 1 | 1 |
| Water | 6 | 7 | 7 | 8 | 1 | 1 |
| Methanol | 5 | 6 | 6 | 7 | 3 | 3 |
| Molecular iodine | 5 | 6 | 5 | 6 | 2 | 3 |
| Carbon dioxide / nitrogen | 4 | 5 | 4 | 5 | 1 | 2 |
| Carbon dioxide / methane | 4 | 5 | 4 | 5 | 2 | 2 |
| Ethylene / ethane | 4 | 5 | 4 | 5 | 2 | 3 |
| Ethanol / water | 7 | 8 | 6 | 8 | 3 | 4 |
| para-Xylene / ortho-xylene | 3 | 3 | 5 | 6 | 2 | 3 |

## Table S11. Common broad-family distribution across the three runs

Table S11. Proposal counts assigned to each of the 30 broad mechanism families in the common comparative taxonomy. Family IDs are arbitrary and carry no ranking. Counts are raw proposal counts; the proposal denominators are 57 for the original structured run, 63 for the repeat structured run and 80 for guest-only.

| Family | Mechanism | Original | Repeat | Guest-only |
|---|---|---|---|---|
| F02 | Adsorption-selective cooperative framework-state switching | 5 | 6 | 28 |
| F01 | Barrier-delayed contraction with excess-inventory expulsion | 4 | 3 | 25 |
| F03 | Guest-biased local conformational control of transport | 8 | 8 | 0 |
| F04 | Direct molecular occlusion of controlling access routes | 7 | 9 | 0 |
| F07 | Resonant guest–framework vibrational hybridization | 7 | 9 | 0 |
| F05 | Collective guest-configurational obstruction | 6 | 8 | 0 |
| F06 | Delayed molecular-hydrogen spin-conversion heat | 1 | 1 | 8 |
| F16 | Cooperative guest orientation–host distortion feedback | 0 | 0 | 9 |
| F23 | Cooperative framework-to-iodine neutral–ionic switching | 0 | 0 | 7 |
| F08 | Quantum-confinement reversal of equilibrium preference | 3 | 3 | 0 |
| F12 | No-overtaking collective transport | 2 | 2 | 0 |
| F13 | Wet-neck-protected internal cavitation | 2 | 2 | 0 |
| F18 | Geometric competition between surface anchoring and guest networks | 2 | 1 | 0 |
| F09 | Tunneling-supported intercavity hydrogen exchange | 1 | 1 | 0 |
| F15 | Water-mediated coupling of local linker conformations | 1 | 1 | 0 |
| F17 | Guest-mediated orientational ordering in an unchanged host | 0 | 1 | 1 |
| F19 | Network-mediated cooperative solvent replacement | 1 | 1 | 0 |

| Family | Mechanism | Original | Repeat | Guest-only |
|---|---|---|---|---|
| F22 | Bridge-controlled water-network percolation | 0 | 2 | 0 |
| F24 | Donor-balanced iodine ionization | 1 | 1 | 0 |
| F28 | Mixed-guest stiffness-disorder phonon scattering | 1 | 1 | 0 |
| F10 | Zero-point-induced entrance exclusion | 1 | 0 | 0 |
| F11 | Spin-isomer-selective entrance memory | 0 | 1 | 0 |
| F14 | Neck-controlled vapor-front release | 1 | 0 | 0 |
| F20 | Confinement-induced dense adsorbate demixing | 0 | 0 | 1 |
| F21 | Water-bridged reorientation of retained ethanol | 1 | 0 | 0 |
| F25 | Charge-transfer-network electronic percolation | 0 | 0 | 1 |
| F26 | Hydrogen-bond relaxation damping of heat-carrying modes | 1 | 0 | 0 |
| F27 | Dynamic elastic bracing by connected water | 1 | 0 | 0 |
| F29 | Hydration-amplified aqua/hydroxo ionization | 0 | 1 | 0 |
| F30 | Site-coordinated water recruits secondary ethanol | 0 | 1 | 0 |

## S9. Supplementary references